\documentclass[pdflatex,sn-aps]{sn-jnl}

\usepackage{graphicx}%
\usepackage{multirow}%
\usepackage{amsmath,amssymb,amsfonts}%
\usepackage{amsthm}%
\usepackage{mathrsfs}%
\usepackage[title]{appendix}%
\usepackage{xcolor}%
\usepackage{textcomp}%
\usepackage{manyfoot}%
\usepackage{booktabs}%
\usepackage{algorithm}%
\usepackage{algorithmicx}%
\usepackage{algpseudocode}%
\usepackage{listings}%
\def\psla{\rlap \slash}

\usepackage{soul}

\theoremstyle{thmstyleone}%
\theoremstyle{thmstyletwo}%

\theoremstyle{thmstylethree}%

\begin{document}

\title[Minkowski Space Dynamics and Light-Front Projection]{Minkowski Space Dynamics and Light-Front Projection}


\author[]{\fnm{Wayne} \sur{de Paula}}\email{wayne@ita.br}

\author*[]{\fnm{Tobias} \sur{Frederico}}\email{tobias@ita.br}
\equalcont{These authors contributed equally to this work.}
\affil[]{\orgname{Instituto Tecnol\'ogico de Aeron\'autica}, \orgaddress{\postcode{12.228-900},\city{ S\~ao Jos\'e dos Campos}, \state{SP}, \country{Brazil}}}


\abstract{We explore the {connection between the  four-dimensional Minkowski-space Bethe–Salpeter equation and its light-front projection, emphasizing the implications for bound-state dynamics. Our approach incorporates dressed  particles}, such as quarks, via the integral representation of the corresponding propagator. We analyze the light-front dynamics of the { valence component of the physical state} using a hierarchical set of Green’s functions, which reveals {  its} coupling to higher Fock {  components} when dressed particles are considered. We also present the light-front Faddeev–Bethe–Salpeter equations for three-body systems with dressed constituents. Furthermore, we discuss formal developments that are central to connecting the { three-dimensional light-front dynamics onto the null-plane and the four-dimensional Minkowski-space framework}, based on the Nakanishi integral representation. Selected applications to hadron structure are also reviewed.}




\maketitle

\section{Introduction}
Bound states in quantum field theory (QFT), particularly in the context of Quantum Chromodynamics (QCD), continue to present deep conceptual and computational challenges. {These difficulties arise due to the need to describe relativistic and non-perturbative phenomena together, such as confinement, dynamical chiral symmetry breaking that leads to the mass generation of the constituent light-quarks within a consistent theoretical framework.} Traditional methods in Euclidean space, like lattice QCD, while highly developed and successful in certain regimes, face limitations when applied to e.g. timelike processes,  that are being successfully pursued in recent years to bridge the gap between space-like and timelike formulations~\cite{Briceno:2017max,Ji:2020ect}.  Minkowski-space formulations of bound-state equations, enables direct access to  physical quantities, albeit at the cost of more intricate analytic structures of the kernel~\cite{Carbonell:2010zw}.

The Bethe-Salpeter equation (BSE)  provides a covariant description of two-body bound states, incorporating full relativistic dynamics and preserving Poincaré symmetry \cite{Salpeter:1951sz}. However, solving the BSE directly in Minkowski space is notoriously difficult due to the presence of singularities and the complex analytic behavior of its solutions. { The development of the Nakanishi integral representation (NIR), originally introduced in the 1960s to represent perturbative Feynman amplitudes \cite{Nakanishi:1971},} has been instrumental in making this problem tractable, reformulating the BSE into an integral equation with a real weight function \cite{Nakanishi:1969ph,Kusaka:1995vv,Carbonell:2006zz,Carbonell:2010zw,Frederico:2011ws,Frederico:2013vga,
Salme:2017oge,Jia:2023imt,Pimentel:2016cpj}.  This representation allows for stable numerical solutions and offers a route to obtain structure observables and in addition access to the light-front (LF) valence wave function, while implicitly taking into account an infinite number of { light-front} Fock-components of the bound state (see e.g.~\cite{Sales:1999ec}). 

Among the three forms of relativistic dynamics analyzed  by Dirac in his seminal paper \cite{Dirac:1949cp}, the one we will focus is the light-front one, which  naturally brings the kinematical variables relevant for high-energy collision processes. The eigenstate of the interacting mass operator, can be defined onto the hypersurface of a fixed light-front time \( x^+ = x^0 + x^3 \), where the relative distances are of space-like nature and therefore without causal connection. This framework provides a Hamiltonian description that has seven kinematical Poincaré generators out of ten,  and therefore the truncation of the Fock space in the light-front Tamm–Dancoff method~\cite{Perry:1990mz} maintains boost invariance~\cite{Brodsky:1997de}. In the context of hadronic physics, light-front wave functions (LFWFs) offer direct links to parton distributions and are natural inputs for scattering processes involving large momentum transfers~\cite{Brodsky:1997de, Bakker:2013cea}.

The intersection of Minkowski-space methods and light-front dynamics yields one practical framework for studying relativistic bound states. In this hybrid approach, the analytic advantages of the NIR and the physical intuition of light-front quantization converge, enabling access to phenomenologically relevant quantities such as electromagnetic form factors, generalized parton distributions (GPDs), and distribution amplitudes \cite{Mezrag:2014jka,dePaula:2016oct}.

Motivating this line of investigation is the growing experimental demand for high-resolution, multidimensional information about the internal structure of hadrons. Next-generation facilities such as the Electron-Ion Collider (EIC) aim to explore the three-dimensional spin and momentum structure of protons and nuclei with unprecedented precision~\cite{Accardi:2012qut,AbdulKhalek:2021gbh,Accardi:2023chb}. These initiatives build upon previous programs at Jefferson Lab, COMPASS, and HERA, which provided the first glimpses into the intricate interplay between quarks and gluons inside hadrons~\cite{Dudek:2012vr, Burkert:2023wzr}. Theoretical efforts must evolve in parallel to meet the interpretive demands of these experiments, particularly in constructing frameworks that can connect experimental observables to the underlying QCD dynamics.

A major theoretical obstacle in constructing such frameworks is Haag’s theorem, which states that the interaction picture is ill-defined in QFT due to the non-unitary equivalence between the free and interacting Hilbert spaces \cite{Haag:1955ev} (see also \cite{Polyzou:2021qpr}). In practice, this implies that field operators in interacting theories cannot be expressed as simple perturbations of free fields, especially in gauge theories like QCD. However, by working with { dressed particles}, we can construct probability amplitudes in practice—an approach that, in our view, is particularly relevant in light-front dynamics~\cite{Brodsky:1997de}, where, for example, dynamical chiral symmetry breaking can be properly incorporated. 

The developments in Minkowski-space solutions to the BSE and their light-front projections, presenting a unified approach to relativistic bound-state dynamics are reviewed. In addition, we extend the methods involving the light-front projection of the BSE and the formulation in Minkowski space through  the Nakanishi representation with { dressed particles}, with an eye toward constructing frameworks consistent with dynamical chiral symmetry breaking. 

The work is organized as follows. In Sect.~\ref{sec:BSE-LFP} we review the BSE and its Light-Front Projection using the quasi-potential approach and now  including { dressed particles}  for bound states of two dressed particles. In Sect.~\ref{sec:hierarchy} the hierarchy equations for the valence LF Green’s functions  with dressed degress of freedom is discussed. In Sect. \ref{sect:FBSLF} it is resented the LF projection of the Faddeev-Bethe-Salpeter equation with dressed particles. In Sect.~\ref{sect: MinkBSE} the Minkowski space BSE and the Nakanishi integral representation is discussed, as well as the relation to the quasi-potential approach for the effective dynamics of the valence component of the wave function, where the inverse Stieltjes transform bridges the two formulations. In Sect.~\ref{sec:hadron} some applications to explore the structure of light hadrons with the solutions of the BSE in Minkowski space are reviewed. The conclusions and outlook of some directions for future work are briefly discussed in Sect. \ref{sec:conclusion}

\section{The Bethe-Salpeter Equation and its Light-Front Projection}\label{sec:BSE-LFP}

The BSE provides a non-perturbative framework for describing relativistic bound states in quantum field theory (QFT) \cite{Salpeter:1951sz}. Its applications range from mesons in QCD to nuclear few-body systems and even to exotic hadrons beyond the quark model. However, the BSE is formulated in Minkowski space, where the interaction kernel and Green's functions possess singular structures that complicate both analytical and numerical treatments. Addressing these complications requires transforming the BSE into a representation where such singularities are manageable, keeping the connection to physical observables.

A promising strategy for overcoming these challenges is the use of the Nakanishi Integral Representation (NIR) \cite{Nakanishi:1969ph, Carbonell:2010zw}. This representation enables the transformation of the four-dimensional integral equation into a two-dimensional one for the weight function. Crucially, this allows us to perform the loop integration, required in Minkowski space, by standard techniques facilitating direct numerical solution of the BSE without recourse to the Euclidean space formulation using Wick rotation. { The NIR maintains full covariance and enables access to timelike quantities such as decay widths and timelike form factors. }

In this context, light-front projection becomes especially valuable. Light-front dynamics (LFD) is formulated on a hypersurface defined by fixed light-front time, \( x^+ = x^0 + x^3 \), leading to a Hamiltonian evolution in this variable. Projection onto the light-front frame effectively reduces the dimensionality of the BSE and simplifies its kinematical structure. This is particularly true for the calculation of wave functions and parton distributions, which naturally arise in the light-front framework \cite{Brodsky:1997de, Brodsky:2000ii,Sales:1999ec,Sales:2001gk}.

The light-front projection of the BSE proceeds by integrating out the relative light-front energy variable \( k^- = k^0 - k^3 \). This yields a light-front wave function (LFWF) that captures the momentum distribution of the bound state constituents in terms of longitudinal momentum fractions and transverse momenta. In the present approach, the projection is carried out using the Nakanishi representation, which permits performing this integration analytically, resulting in an expression involving only the weight function and the light-front kinematics \cite{dePaula:2016oct,dePaula:2017ikc, Carbonell:2017kqa}.

This method preserves the manifest boost invariance in the longitudinal direction and aligns naturally with partonic descriptions of hadrons. In particular, LFWFs derived from the BSE can be directly related to parton distribution amplitudes (PDAs), generalized parton distributions (GPDs), and transverse momentum dependent distributions (TMDs). These observables are central to modern experimental programs, such as those at Jefferson Lab and the upcoming Electron-Ion Collider (EIC)~\cite{Accardi:2023chb, AbdulKhalek:2021gbh}.

The relation between light-front wave functions (LFWFs) and experimentally measurable quantities has renewed interest in applying the BSE formalism in light-front coordinates. This connection is essential for extracting generalized parton distributions and transverse structure functions, providing a bridge between theory and processes such as deeply virtual Compton scattering (DVCS), meson electroproduction, and jet physics. Access to timelike data further enables the extraction of crucial information necessary for constructing a comprehensive picture of QCD in both the valence and sea regions \cite{Diehl:2003ny,Belitsky:2005qn}.

From a computational standpoint, the combination of the BSE in Minkowski space with light-front projection reduces the complexity of the problem while retaining access to physical observables. Recent advances in numerical methods, such as the use of basis light-front quantization (BLFQ) \cite{Vary:2009gt,Li:2015zda} and the expansion of the Nakanishi weight function in orthogonal polynomials, have made these calculations more tractable~\cite{dePaula:2017ikc}. Furthermore, this framework allows the consistent inclusion of dressed propagators and vertices, which are necessary for capturing non-perturbative QCD dynamics and for preserving symmetries such as gauge invariance and for dynamical chiral symmetry breaking~\cite{Binosi:2014aea,Roberts:2016vyn}.{ We add that within BLFQ spontaneous chiral symmetry breaking has already been handled  with color singlet Nambu-Jona-Lasinio interactions~\cite{Jia:2018ary}.
}

An important benefit of the light-front projection is its ability to address certain formal issues in QFT. Notably, Haag’s theorem implies that the interaction picture does not exist in a fully covariant formulation of QFT due to the unitary inequivalence of free and interacting representations. However, light-front dynamics offers a practical solution by employing a different quantization surface, where vacuum fluctuations are suppressed and the Fock expansion of physical states becomes meaningful \cite{Brodsky:1997de, Haag:1955ev}. However, this conceptual simplification leaves out the LF zero-modes that has severe implications for symmetry breaking \cite{Tsujimaru:1997jt} challenging the notion of a meaningful LF Fock-space expansion~\cite{Nakanishi:1976vf}. Later on, in Sect.~\ref{lfval} we will return to this relevant discussion.

Recent advances have enabled direct solutions of the BSE for two-fermion bound states in Minkowski space \cite{dePaula:2016oct,dePaula:2017ikc}. These developments rely on the Nakanishi integral representation of the Bethe-Salpeter amplitude, combined with light-front momentum parametrization. A key breakthrough lies in the exact analytical treatment of singularities, which previously hindered progress, allowing the method to accommodate complex interaction kernels. This framework opens the door to realistic, non-perturbative studies of relativistic bound systems with arbitrary spin. Comparative analyses with results obtained in Euclidean and Minkowski formulations confirm the robustness of the approach. Additionally, newly derived light-front amplitudes illustrate its practical potential for phenomenological applications.

Finally, the BSE in combination with light-front techniques has also proven effective in exploring exotic hadronic configurations such as tetraquarks, glueballs, and hybrid mesons. These states often involve strong coupling and multi-particle dynamics, where traditional methods struggle. The light-front projected BSE can accommodate these features by allowing the systematic inclusion of higher Fock components and providing access to their distributions and decay modes \cite{Eichmann:2015nra,Sanchis-Alepuz:2015qra,Lebed:2016hpi}.

In summary, the BSE, when reformulated using the Nakanishi integral representation and light-front projection, yields a powerful and versatile tool for probing the internal structure of hadrons by performing calculations of their structure functions in Minkowski space \cite{dePaula:2020qna,dePaula:2023ver,dePaula:2022pcb,Ydrefors:2021dwa}. This hybrid approach combines the strengths of covariant field theory and light-front quantization, making it ideally suited for tackling the non-perturbative frontier of QCD and for interpreting the results of modern experimental programs aimed at mapping the partonic landscape of hadrons. The light-front projected BSE, enriched by renormalization group invariance and consistent subtraction schemes, stands as a promising path toward understanding hadron structure from first principles. { We have to add that, the LF projection of the BS amplitude allows to access the valence component of the light-front wave function, while the higher Fock-components are implicitly taken into account in the kernel of the LF projected BS equation, and in any observable computed through the BS amplitude, while the light-front dynamics is able to get, like in BLFQ, explicitly the contribution of each Fock-component to a given observable (see the recent work~\cite{Lan:2024ais}).}

\subsection{Minkowski space BSE's} \label{QPred}

The BSE (see, e.g., \cite{Gross:1993zj}) in the four-dimensional Minkowski space provides the fundamental starting point for many investigations of few-body systems incorporating relativistic effects. It is formulated within a fully covariant quantum field-theoretical framework based on an interacting Lagrangian that models the system under study.

The Minkowski space Green's function for two particles, or the connected four-point function, can be written in terms of the transition matrix $T(K)$, with total four-momentum $K$, as: \begin{equation} \mathcal{G}(K)=\mathcal{G}_0(K)+\mathcal{G}_0(K)\,T(K)\,\mathcal{G}_0(K)\,, \end{equation} where the disconnected two-body Green’s function $\mathcal{G}_0(K)$ should, in general, include self-energy corrections. The Fourier transform of the connected Green's function provides the amplitude for interacting particles that propagate from some initial individual space-time positions to final ones. The interpretation as a probability amplitude emerges when the initial and final space-time separations between the particles are space-like.

The transition matrix $T(K)$ satisfies a 4D equation that reads \begin{eqnarray} T(K) = V(K) + V(K)\mathcal{G}_0(K)T(K)\,, \label{1} \end{eqnarray} where $V(K)$ contains, in principle, all two-body irreducible contributions. The solution of Eq.(\ref{1}) yields the $T$-matrix, which encodes the full dynamics of both scattering and bound-state processes. However, in previous applications of the quasi-potential (QP) approach\cite{Sales:1999ec,Sales:2001gk}, which performed the three-dimensional reduction to the LF, self-energy contributions were neglected for simplicity; these will now be addressed in the formal developments that follow.

The dressing of the particles in the disconnected Green’s function for two bosons takes the form \begin{equation} \mathcal{G}_0(K) = \frac{\imath^2}{2\pi} \int_0^\infty \mathrm{d}s_1^2~ \frac{\rho(s_1^2)}{\widehat{k}_1^2 - s_1^2 + i\epsilon} \int_0^\infty \mathrm{d}s_2^2~\frac{\rho(s_2^2)}{\widehat{k}_2^2 - s_2^2 + i\epsilon}\,, \label{2a} \end{equation} where the Källén–Lehmann (KL) spectral representation is employed to account for the dressing of the individual propagators. Here, $\widehat{k}_i$ denotes the four-momentum operator, and the factor $2\pi$ is introduced for convenience.

In the fermionic case, using the KL representation, the disconnected Green’s function becomes \begin{equation} \mathcal{G}_0(K) = \frac{\imath^2}{2\pi} \int_0^\infty \mathrm{d}s_1^2~\frac{\rho_F(s_1^2,\widehat{k}_1)}{\widehat{k}_1^2 - s_1^2 + i\epsilon} \otimes \int_0^\infty \mathrm{d}s_2^2~ \frac{\rho_F(s_2^2,\widehat{k}_2)}{\widehat{k}_2^2 - s_2^2 + i\epsilon}\,, \label{2aF} \end{equation} where the fermionic spectral density is given by $\rho_F(s^2) = \rho^V(s^2)\, \widehat{\psla k} + \rho^S(s^2)$, which incorporates both the vector ($V$) and scalar ($S$) components of the fermion two-point function.

The two-particle bound state, characterized by a total four-momentum $K_B$ satisfying $K_B^2 = M_B^2$, corresponds to a pole of the $T$-matrix. The residue at the pole is associated with the vertex function, which identifies the nontrivial part of the Bethe-Salpeter (BS) amplitude $\Psi_B$, satisfying the homogeneous equation \begin{eqnarray} \left| \Psi_B \right\rangle = \mathcal{G}_0(K_B)V(K_B)\left| \Psi_B \right\rangle\,. \label{1.2ab} \end{eqnarray} The amplitude $\left| \Psi_B \right\rangle$ must additionally fulfill a proper normalization condition~\cite{Gross:1993zj}.

For scattering states with total momentum $K$, the BS amplitude satisfies the inhomogeneous equation \begin{eqnarray} \left| \Psi^+ \right\rangle = \left| \Psi_0 \right\rangle + \mathcal{G}_0(K)V(K)\left| \Psi^+ \right\rangle\,, \label{1.2as} \end{eqnarray} which provides the scattering amplitude for the two-particle system.

The quasi-potential reduction: a tool for eliminating the relative LF-time is introduced in what follows, and the basic operator is the LF projected disconnected Green's function, that we are going to present for two-boson and two-fermion systems.

\subsection{Disconnected global LF two-boson propagator} \label{sec:global2boson}

Since in the LF projection method a central role is played by the on-minus-shell propagation, as it
will be emphasized below, let us write the relevant matrix elements of the free two-boson
Green's function, viz
\begin{multline}
\left\langle k_1^{\prime -}\right| G_0(K)\left|k^-_1\right\rangle
= \frac{\imath^2}{2\pi}\frac{\delta \left(k^{\prime -}_1-k^{-}_1\right)}{\widehat k
^+_1 (K^+-\widehat k^+_1)}\\ \times\int_0^\infty \hspace{-.2cm}ds_1^2\int_0^\infty \hspace{-.2cm}ds_2^2
\frac{\rho\big(s_1^2\big)~\rho\big(s_2^2\big)}{ \left(k_1^--\widehat k_{1on}^{s-}+\frac{i\varepsilon}{\widehat k^+_1}\right)
\left(K^--k_1^--\widehat k_{2on}^{s-}+\frac{ i\varepsilon}{K^+-\widehat k^+_1} \right)}~,
\label{3}
\end{multline}
where  the LF  four-momenta are  $k_i=(k^-_i:=k^0_i-k^3_i\ ,
\ k^+_i:=k^0_i+k^3_i \ , \ \vec k_{i\perp})$, $
k^{s-}_{ion}=(\vec k_{i\perp}^2+s_i^2)/ k^+_i$
($i=1,2$) is the on-minus-shell
momentum operator, with eigenfunctions given by  the LF plane waves,   $ \langle x^-_i \vec
x_{i\perp}\left| k^+_i\vec k_{i\perp}\right\rangle= {\cal N}~
e^{-\imath(\frac12 k^+_ix^-_i-\vec k_{i\perp} . \vec x_{i\perp})}$.
 The  completeness relation and the normalization are
\begin{eqnarray} \int \frac{dk^+d^2k_\perp}{2(2\pi)^3} \left. |k^+\vec
k_\perp\rangle \langle k^+\vec k _\perp \right. |
=\mbox{\boldmath$1$}, \label{nboson}\end{eqnarray} and $\langle
k^{\prime +}\vec k^\prime_\perp |k^+\vec k_\perp
\rangle=2(2\pi)^3\delta(k^{\prime +}-k^+)\delta(\vec
k^{\prime}_\perp-\vec k_\perp)  $, respectively.

The first step for projecting the BSE onto the LF surface is the introduction
of
the free-resolvent, i.e., the Fourier transform in $K^-$ of the
global $x^+$-time free propagator of the two-particle system. This  amounts
to  integrate the matrix elements, Eq. (\ref{3}) (or Eq. (\ref{3b})), of the 4D $G_0(K)$
over $k^-_1$ and $k^{\prime-}_1$, so that   the relative LF
time between the particles is eliminated, and one remains with a dependence upon $K^-$, i.e.
\begin{eqnarray}
| G_0(K)|:=  \int dk^{\prime -}_1 dk^{ -}_1
\left\langle
k_1^{\prime -}\right|  G_0(K)\left|k^-_1\right\rangle\equiv g_0(K)
\label{2.11a}
\end{eqnarray}
where $g_0(K)$ is called the free-global LF propagator. It is a 3D operator depending upon the LF
momenta $(k^+_i,\vec k_{i\perp })$ only. In  the above equation the vertical bars `` $|$ ''on the right side and  on
the left one indicate that the minus components  in $|k^-\rangle$
and $\langle k^{\prime -}|$ have to be integrated
out \cite{Sales:1999ec,Sales:2001gk}, respectively. 

Evaluating the integrals in Eq.~\eqref{2.11a} explicitly given by
\begin{eqnarray}
g_0(K)= i\,\frac{\theta (K^{+}-\widehat{k}_{1}^{+})\theta (\widehat{k}_{1}^{+})}{ \widehat k^+_1 (K^+-\widehat k^+_1)} \int_0^\infty ds_1^2\int_0^\infty ds_2^2 \,\frac{\rho(s_1^2)\,\rho(s_2^2)}{K^{-}-\widehat{%
k}_{1on}^{-}-\widehat{k}_{2on}^{-}+i\varepsilon} \ ,
\label{2.11}
\end{eqnarray}
where positive value for
 $K^+$  was considered without any loss of generality.

\subsection{Disconnected global LF two-fermion propagator}\label{sec:global2fermion}

In order to introduce the two-fermion global propagator, we have to separate the non-propagating part of the Dirac propagator, or the instantaneous LF term  as follows (see Ref.~\cite{Yan:1973qg}):
\begin{equation}
\int_0^\infty \hspace{-.15cm}ds^2 ~\frac{\rho_F(s^2,k) }{k^2-s^2+i\epsilon}=\hspace{-.1cm}\int_0^\infty\hspace{-.15cm} ds^2 ~\frac{\rho_F(s^2,{k}_{on}) }{k^+(k^--k^{s-}_{on}+ \frac{i\varepsilon}{
k^+})}+\frac{\gamma^+}{2k^+}\int_0^\infty \hspace{-.15cm}ds^2 \,\rho_V(s^2)\,,\label{instant}
\end{equation}
where 
$\rho_F(s^2,k)=\rho_V(s^2)\psla{k}+\rho_S(s^2)\,, $
and  the first term of Eq.~\eqref{instant} yields the on-minus-shell propagation, while the second one the LF-time instantaneous term
of the Dirac propagator.  One may wonder what happens  with the instantaneous term in QCD,  namely, if it could vanish. However, asymptotic freedom says that the quark propagator in the ultraviolet should approach the free one, and therefore
\begin{equation}\label{eq:sumrule}
 \int_0^\infty \hspace{-.15cm}ds^2 \,\rho_V(s^2)=1 \quad\text{and}\quad   \int_0^\infty \hspace{-.15cm}ds^2 \,\rho_S(s^2)=m_0\,, 
\end{equation}
where $m_0$ is the current quark mass. Thus, asymptotic freedom does not allow to ignore the instantaneous term in the quark propagator.

The two-fermion disconnected Green's function written in terms of the LF momentum operators is:
\begin{multline}
\left\langle k_1^{\prime -}\right| G^F_0(K)\left|k^-_1\right\rangle
= \frac{\imath^2}{2\pi}\frac{\delta \left(k^{\prime -}_1-k^{-}_1\right)}{\widehat k
^+_1 (K^+-\widehat k^+_1)}\\ \times\int_0^\infty \hspace{-.2cm}ds_1^2\int_0^\infty \hspace{-.2cm}ds_2^2
\frac{\rho_F\big(s_1^2,k_1\big)\otimes\rho_F\big(s_2^2,k_2\big)}{ \left(k_1^--\widehat k_{1on}^{s-}+\frac{i\varepsilon}{\widehat k^+_1}\right)
\left(K^--k_1^--\widehat k_{2on}^{s-}+\frac{ i\varepsilon}{K^+-\widehat k^+_1} \right)}~,
\label{3F}
\end{multline}
where for the sake of our  notation $\widehat k_i^\mu=\{k^-_i,\widehat k^+_i,\widehat {\vec k}_{i\perp}\}$. The LF propagating part that allows to build  the  analogous of the bosonic Eq.~\eqref{3}) is written as:
\begin{multline}
\left\langle k_1^{\prime -}\right| G_0(K)\left|k^-_1\right\rangle
= 
\frac{\imath^2}{2\pi}\frac{\delta \left(k^{\prime -}_1-k^{-}_1\right)}{\widehat k^+_1 (K^+-\widehat k^+_1)}
\\ \times\int_0^\infty \hspace{-.2cm}ds_1^2\int_0^\infty \hspace{-.2cm}ds_2^2
\frac{ \rho_F\big(s_1^2,\,\widehat k_{1on}\big)\,  \otimes\,\rho_F\big(s_2^2,\,\widehat k_{2on}\big)\, 
}{ \left(k_1^--\widehat k_{1on}^{s-}+\frac{i\varepsilon}{\widehat k^+_1}\right)
\left(K^--k_1^--\widehat k_{2on}^{s-}+\frac{ i\varepsilon}{K^+-\widehat k^+_1} \right)}~,
\label{3a}
\end{multline}
and performing the projection onto the light-front the resolvent for the dressed two-fermion system becomes:
\begin{equation}
 |G_0(K)|=i\,\frac{\theta (K^{+}-\widehat{k}_{1}^{+})\theta (\widehat{k}_{1}^{+})}{ \widehat k^+_1 (K^+-\widehat k^+_1)} \int_0^\infty \hspace{-.2cm}ds_1^2\int_0^\infty \hspace{-.2cm}ds_2^2\,\,
\frac{\rho_F\big(s_1^2,\,\widehat k_{1on}\big)\,  \otimes\,\rho_F\big(s_2^2,\,\widehat k_{2on}\big)\, 
}{ 
K^--\widehat k_{1on}^{s-}-\widehat k_{2on}^{s-}+ i\varepsilon}~,
\end{equation}

In order to build the global LF propagator we introduce the spinor projector on the positive energy states with self-energy $\Sigma_i(k^+,k_\perp)$, to be defined in the frame where the total transverse momentum vanishes:
\begin{equation}\label{eq:dressproj}
 \Lambda^+(\widehat k^\Sigma_{ion})=\frac{\rlap\slash
\widehat{k}^\Sigma_{ion} +\Sigma_i(\widehat k_i^+,\widehat{\vec k}_{i\perp})}{2\,\Sigma_i(\widehat k_i^+,\widehat{\vec k}_{i\perp})}\,,
\end{equation}
with the on-minus-shell four-momentum is
$k^\Sigma_i=(k^{\Sigma-}_i\ ,
\ k^+_i \ , \ \vec k_{i\perp})$ and $
k^{\Sigma-}_{ion}=({\vec k}_{i\perp}^2+\Sigma_i(k^+_i,\vec k_{i\perp})^2)/ \widehat k^+_i$ ($i=1,2$). The light-front fermion self-energy is essential to take into account the fermion dressing to define the global LF two-fermion propagator. It trivially reduces to the standard LF spinor projector when the the self-energy corresponds to a fixed constituent mass. The construction of the LF self-energy, although quite interesting, as it defines the dressed state, and in QCD the dressed quark which embodies the complex QCD dynamics leading to chiral symmetry breaking and the nucleon mass (see e.g.~\cite{Aguilar:2019teb,Roberts:2021nhw,AbdulKhalek:2021gbh}) is postponed to a future work.

The global LF  two-fermion propagator is then written as:
\begin{equation}
g_0(K)
= i\,\frac{\theta (K^{+}-\widehat{k}_{1}^{+})\theta (\widehat{k}_{1}^{+})}{ \widehat k^+_1 (K^+-\widehat k^+_1)} \int_0^\infty \hspace{-.2cm}ds_1^2\int_0^\infty \hspace{-.2cm}ds_2^2\,\,
\frac{ \widehat A^+_1\otimes \widehat A^+_2 
}{
K^--\widehat k_{1on}^{s-}-\widehat k_{2on}^{s-}+i\varepsilon}~\,,
\label{3b}
\end{equation}
where the Dirac operator is 
\begin{equation}
\widehat A^+_i=\,\Lambda^+\big(\widehat {k}^\Sigma_{ion}\big)\,\rho\big(s^2,\,\widehat k_{ion}\big)\,\Lambda^+\big(\widehat {k}^\Sigma_{ion}\big)\,.
\label{eq:Aplus}
\end{equation}
Elaborating the  operator defined in Eq.~\eqref{eq:Aplus}, we find that
\begin{eqnarray}
\widehat A^+_i
&=&\mathcal{D}_s(\widehat 
 y_i)\,\Lambda^+\big(\widehat {k}^\Sigma_{ion}\big) \quad \text{where}\quad \widehat  y_i=\frac{\widehat k^{\Sigma}_{ion}\cdot \widehat k^s_{ion}}{ \Sigma_i(\widehat k_i^+,\widehat{\vec k}_{i\perp}) }\,,
\end{eqnarray}
and
$\mathcal{D}_s(y ) = \rho^S(s)+ \,\rho^V(s) \,y\,$.
From those above equations, we rewrite the projected global free propagator for dressed two-fermion systems as:
\begin{equation}
g_0(K) = \left( \widehat{\rlap\slash
k}_{1on} +\Sigma_1(\widehat k^+_1,\widehat{\vec k}_{1\perp})\right) \otimes\left(\widehat{\rlap\slash k}_{2on} +
\Sigma_2(\widehat k^+_2,\widehat{\vec k}_{2\perp})\right) 
\overline g_{0}(K) \, ,
\label{3b1}    
\end{equation}
where
\begin{equation}
\overline g_{0}(K)
=i\,\frac{\theta (K^{+}-\widehat{k}_{1}^{+})\theta (\widehat{k}_{1}^{+})}{ \widehat k^+_1 (K^+-\widehat k^+_1)}\int_0^\infty \hspace{-.2cm}ds_1^2\int_0^\infty \hspace{-.2cm}ds_2^2\,\,
{ \,{\mathcal{D}_1(y_1)\over 
2\,\Sigma_1(\widehat k_1^+,\widehat{\vec k}_{1\perp})}\,\,{\mathcal{D}_2(y_2)\over 2\,\Sigma_2(\widehat k_2^+,\widehat{\vec k}_{2\perp})}
\over 
K^--\widehat k_{1on}^{s-}-\widehat k_{2on}^{s-}+i\varepsilon }~\,,
\label{3F1}
\end{equation}
which is the generalization to the two-fermion case of Eq.~\eqref{2.11} for the two-boson global LF propagator. The inverse of $g_0(K)$ exists only in the valence
sector, satisfying the theta functions, since the projectors, $\Lambda_+$, single out only positive energy states. For the moment the self-energy $\Sigma(k^+,k_\perp)$ is instrumental and will be not tackled in what follows. We observe the present analysis can be extended to systems composed by
particle-antiparticle or by other mixtures, like  a fermion and a boson~\cite{AlvarengaNogueira:2019zcs}.

If one considers  QCD in the ultraviolet (UV) limit the running quark mass should approach the current one, and taking into account the sum rules~\eqref{eq:sumrule}, the  global LF two-quark propagator becomes the free one. In the phenomenological case of a fixed constituent quark mass,  the global LF propagator reduces to the standard one~\cite{Sales:2001gk}, once the spectral densities are substituted by 
$$\rho^S(s^2)=m\,\delta(s^2-m^2)\quad\text{and}\quad \rho^V(s^2)=\delta(s^2-m^2)\,, $$
which leads to:
\begin{equation}
  g_0(K)
= 
{ (\rlap\slash
\widehat{k}_{1on} +m) \otimes   (\rlap\slash
\widehat{k}_{2on} +m)
\over \widehat k^+_1 (K^+-\widehat k^+_1)
(K^--\widehat k_{1on}^--\widehat k_{2on}^-+i\varepsilon)}~\,,
\label{3bb}
\end{equation}
where $k^-_{ion}=(\vec k^2_\perp+m^2)/k^+_i$. with $m$ being the constituent mass.

\subsection{The QP Method and LF Projection}\label{sec:QPMLFProj}

In the QP formalism \cite{wolja}, an auxiliary interaction, $W(K)$, is introduced to allow the three-dimensional reduction, leading to the following system of coupled equations for the full T-matrix:
\begin{eqnarray} && T(K) = W(K) + W(K)\widetilde{G}_{0}(K)T(K), \label{2.1} \\ && 
W(K) = V(K) + V(K)\Delta_0(K)W(K), \label{2.3a} \end{eqnarray} where for bosons one has that $\Delta_0(K) = G_{0}(K) - \widetilde{G}_{0}(K)$ and  for fermions $\Delta_0(K) = G^F_{0}(K) - \widetilde{G}_{0}(K)$. The auxiliary Green's function $\widetilde{G}_0(K)$ is a four-dimensional operator that depends on the four-momenta of the system's constituents and plays an essential role in the LF projection. It can be thought of as the four-dimensional counterpart of the three-dimensional Green's function $g_0(K)$, which notably does not include the relative-time propagation of the system. As such, $\Delta_0(K)$ accounts for this propagation in four-dimensional space, making it a key element to recover the Minkowski space internal dynamics of the system.

The auxiliary 4D Green's function $\widetilde{G}_0(K)$ is defined as: \begin{equation} 
\widetilde{G}_0(K) = \bar{\Pi}_0(K) g_0(K) \Pi_0(K) \, , \label{g0tilde} 
\end{equation} 
where \begin{eqnarray} \bar{\Pi}_0(K) = G_0(K) |~g^{-1}_0(K), \quad\text{and}\quad \Pi_0(K) =g^{-1}_0(K) | G_0(K). \end{eqnarray} 
These operators, termed as the free reverse LF-projection operators in Ref.~\cite{Frederico:2010zh}, serve as the bridge between three-dimensional and four-dimensional quantities, and the  essential ingredients are the free LF global propagators Eq.~\eqref{2.11} for dressed two-bosons and Eq.~\eqref{3b} for dressed two-fermions systems.

In the subsequent section, we will introduce the corresponding interacting operators. The form of $\widetilde{G}_0(K)$ and the associated integral equation for $W(K)$ (Eq. (\ref{2.3a})) permit the inclusion of only LF two-body irreducible terms.

The iterative solution of Eq. (\ref{2.3a}) is expressed as \begin{eqnarray} W(K) = \sum_{n=1}^\infty W_n(K), \label{expw} \end{eqnarray} where $W_n(K) = V(K) (\Delta_0(K) V(K))^{n-1}$. A diagrammatic analysis of the series (\ref{expw}) reveals that each term corresponds to a distinct Fock content, related to the propagation of virtual intermediate states, as detailed in Section \ref{sec:hierarchy}. {  It is worthwhile to remark that by truncating the sum in Eq. (\ref{expw}), namely, the number of Fock components considered in the calculation, the QP method gives an approximate solution of the BS equation.
}

For fermionic models with point-like interactions, the LF BSE was studied by retaining the lowest-order kernel, revealing subtle issues related to divergences from transverse momentum dependence \cite{Glazek:1992bs,Mangin-Brinet:2001wmj,ManginBrinet:2003nm}. These difficulties are partly attributed to the absence of instantaneous terms, as discussed in \cite{Sales:2001gk,Bakker:2006pn}.

Following \cite{Sales:1999ec,Sales:2001gk,Marinho:2007zzb,Marinho:2008pe}, it is possible to derive a  LF T-matrix, $t(K)$, from the 4D T-matrix. Specifically, we have: \begin{equation} t(K) = \Pi_0(K)T(K)\overline{\Pi}_0(K) = w(K) + w(K)g_0(K)t(K) \,, \label{3.1} \end{equation} where the LF 3D effective interaction is $w(K) := \Pi_0(K)W(K)\overline{\Pi}_0(K)$, which takes into account the coupling between different Fock components. 
 
 The  interacting LF Green's function, $g(K)$, satisfies the integral equation: \begin{equation} g(K) = g_0(K) + g_0(K)w(K)g(K)\,, \label{LFRESOLV0} \end{equation}
 as in nonrelativistic scattering theory it also holds that $g(k) = g_0(k) + g_0(k)t(K)g_0(k)$ and  $t(K) = w(K) + w(K)g(K)w(K)$. The operator $g(K)$ is the Fourier transform in $K^-$ of the global LF-time propagator: \begin{eqnarray} g(K) = | G_0(K) | + | G_0(K)T(K)G_0(K) |\,, \label{greenproj} \end{eqnarray} and it evolves the  two-particle valence { component} from an initial LF hypersurface to another one in the future LF-time. { It has to be emphasized that the valence component of the LF wave function, through its evolution will couple to higher Fock-components, as implicitly taken into account by the operator $| G_0(K)T(K)G_0(K) |$ in Eq.~\eqref{greenproj}, in a way that it is a reminiscent of  the "iterated resolvent method"  discussed in Ref.~\cite{Brodsky:1997de}, and also is closely related to the Basis-Light-Front-Quantization (BLFQ) method~\cite{Vary:2009gt,Vary:2016emi}, when the Tamm–Dancoff set of coupled LF eigenvalue equations for the LF Hamiltonian are reduced to the valence sector (see e.g.~\cite{Lan:2024ais}).}

By iterating Eq. (\ref{2.1}) once and using Eqs. (\ref{g0tilde}) and (\ref{3.1}), we obtain: \begin{eqnarray} 
T(K) &=& W(K) + W(K)\left[\widetilde{G}_0(K) + \widetilde{G}_0(K)T(K)\widetilde{G}_0(K)\right] W(K) \nonumber \\ &=& W(K) + W(K) \bar{\Pi}_0(K) g(K) \Pi_0(K) W(K), \label{3.2} \end{eqnarray} which allows the mapping of the LF dynamics into the Minkowski space.
The on-mass-shell matrix elements of $T(K)$, define the two-constituent scattering amplitude and are identical to those obtained from the on-minus-energy-shell matrix elements of $t(K)$, as discussed in \cite{Sales:1999ec,Sales:2001gk}. The "$+\imath \varepsilon$" prescription takes into account the necessary boundary condition for the forward propagation in the LF-time.

\subsection{The Minkowski BS Amplitude and the LF valence { component}}
\label{lfval}

The  Minkowski space BS amplitude $\left|\Psi_B\right\rangle$ for a bound state with total momentum $K_B$ can be computed directly from the valence component of the LF  wave function $\left|\Phi_B\right\rangle$ ~\cite{Sales:2001gk} as:
\begin{equation}
\left|\Psi_B\right\rangle = \mathcal{G}_0(K_B) W(K_B) G_0(K_B) \big|g_0^{-1}(K_B)\, \left|\Phi_B\right\rangle \ ,
\label{3.6a}
\end{equation}
where the valence component is obtained as the solution of the eigen equation:
\begin{equation}
\left|\Phi_B\right\rangle = g_0(K_B) w(K_B) \left|\Phi_B\right\rangle \ ,
\label{lfbse}
\end{equation}
In the case of fermions, it is important to note that the instantaneous terms from the Dirac propagators contribute to the expressions for i) $\mathcal{G}_0 \equiv G_0^F$, ii) $W$, and iii) the effective interaction $w$.

We can introduce the identity $ G_0(K_B) \left[g_0^{-1}(K_B) - w(K_B)\right] \left|\Phi_B\right\rangle = 0$, which when added to Eq. (\ref{3.6a}) gives:
\begin{equation}
\left|\Psi_B\right\rangle = \left[1 + \Delta_0(K_B) W(K_B)\right] G_0(K_B) \big|g_0^{-1}(K_B)\,\left|\Phi_B\right\rangle \ ,
\label{psicov}
\end{equation}
This relation is valid not only for bound states but also for scattering states (see e.g.~\cite{Marinho:2008pe}), i.e.:
\begin{equation}
\left|\Psi^+\right\rangle = \Pi(K) \left|\Phi^+\right\rangle \ ,
\label{psicov1}
\end{equation}
where $\Pi(K)$, the interacting reverse LF-projection operator, is expressed as:
\begin{equation}
\Pi(K) := \left[1 + \Delta_0(K) W(K)\right] G_0(K) \big|g_0^{-1}(K) = G_R(K) \big| g^{-1}(K) \ .
\label{invop}
\end{equation}
with:
\begin{equation}
G_R(K) := G_0(K) + \mathcal{G}(K) V(K) G_0(K) = \mathcal{G}(K) \mathcal{G}_0^{-1}(K) G_0(K) \ .
\label{GR}
\end{equation}
In $G_R(K)$, the on-shell Green's function $G_0(K)$ appears at the rightmost position, which leads to applying the "$|$" operation on the right.

The operator $\Pi(K)$ acts on the valence component of the LF wave function in Eqs. (\ref{psicov}) and (\ref{psicov1}), and this allows the full reconstruction of the 4D BS amplitude for both bound and scattering states, starting from the valence wave function. The conjugate LF operator, $\bar\Pi(K)$, is given by:
\begin{equation}
\bar\Pi(K) := g^{-1}(K) \big| G_L(K)  \ ,
\end{equation}
with:
\begin{equation}
G_L(K) := G_0(K) + G_0(K) V(K) \mathcal{G}(K) = G_0(K) \mathcal{G}_0^{-1}(K) \mathcal{G}(K) \ .
\label{GL}
\end{equation}
This operator allows the "$|$" operation on the left. Reverse LF projectors compactly represent the relationships between operators and states on a 3D hypersurface and their Minkowski space counterparts. Specifically, the relation between the BS amplitude and the valence component can be written as:
\[
|\Psi\rangle = \Pi(K) |\Phi\rangle
\quad \text{and} \quad
\langle \Psi | = \langle \Phi | \bar \Pi(K) \ .
\]
These relations are applicable to both two-boson~\cite{Sales:1999ec,Sales:2001gk} and two-fermion systems~\cite{Marinho:2007zzb,Marinho:2008pe}, with the appropriate choice of $\mathcal{G}$, $\mathcal{G}_0$, and the on-shell $G_0(K)$. Conversely, the valence component of the LF wave function can be directly obtained from the BS amplitude by using Eqs. (\ref{invop}) and (\ref{GR})~\cite{Sales:1999ec,Sales:2001gk,Marinho:2007zzb,Marinho:2008pe}:
\begin{equation}
| G_0(K) \mathcal{G}_0^{-1}(K) |\Psi\rangle = | G_0(K) \mathcal{G}_0^{-1}(K) \mathcal{G}(K) \mathcal{G}_0^{-1}(K) G_0(K) | g^{-1}(K) |\Phi\rangle = |\Phi\rangle \ .
\label{psiproj}
\end{equation}

The valence light-front wave functions for both bound and scattering states are solutions of the squared-mass eigenvalue equation, namely $ g^{-1}(K) |\Phi\rangle = 0 \,, $ having the appropriate boundary conditions for each case. This eigenvalue equation, reduced to the valence sector, incorporates an effective interaction that accounts for the coupling to higher Fock states, as we have already mentioned, naturally associated with the "iterated resolvent method"~\cite{Brodsky:1997de}. To be complete we recall that the LF scattering state is the solution of the inhomogeneous equation
$|\Phi^+\rangle = |\Phi_0\rangle + g_0(K) w(K) |\Phi^+\rangle\,,$
that incorporates an outgoing boundary condition.

In addition, it is important to recall once more that the successful method of Basis Light-Front Quantization (BLFQ) applied to bound states~\cite{Vary:2009gt,Vary:2016emi} involves solving a set of coupled equations for the eigenvalues of the LF Hamiltonian within a Fock-space decomposition. When this system is reduced to the valence sector, the resulting effective interaction captures the Hamiltonian couplings among different Fock sectors. As an illustrative example, in a truncated Fock-space representation for the pion LF wave function, schematically written as $|\pi\rangle= \psi_{q\bar q}|q\bar q\rangle+\psi_{q\bar qg}|q\bar qg\rangle$, the three-particle amplitude $\psi_{q\bar qg}$ has been derived from the valence component (see, e.g.,~\cite{Lan:2024ais}). Starting from the BSE for two fermion system in the light-cone gauge by 
implementing the Leibbrandt-Mandelstam prescription, the performed projection of the bound state equation to the light-front~\cite{LiuSoperPRD1993} could be also derived by the quasi-potential technique, delineated  in this section, once the bare fermion propagator is used. However, by taking into account the dressed fermion propagator, it departs from the standard Tamm-Dancoff truncation of the light-front Fock space~\cite{Perry:1990mz}, as the fully dressed particles come with an infinite content in the bare Fock-sector. 

 A challenge is posed by Haag's theorem~\cite{Haag:1955ev} that states that the  free and interacting quantum field-theory Hilbert spaces are nonequivalent, contrasting with  the LF quantization based on the Fock-space expansion with eigenstates of the free LF Hamiltonian.  The attempt to overcome this problem came with the formulation of the LF Quantum Field-Theory in Ref.~\cite{Nakanishi1977}, which reached the conclusion that one way to allow a consistent LF quantization is to adopt a prescription based on a infinitesimal parameter to approach the null-plane in the vanishing limit. However, as discussed in~\cite{Nakanishi:1976vf}
the vacuum polarization does not disappear, and  the  physical Fock-space cannot the constructed, challenging the usefulness of the LF
wave function and the triviality of the LF vacuum, which we know now is plagued by zero-modes~\cite{deMelo:1998an,deMelo:2012hj,Collins:2018aqt,Chabysheva:2025vtm}. 
Nevertheless, Ref.~\cite{Ji:2020ect}  suggested an effective theory point of view where the zero-modes are cut off from the LF vacuum and leaving their contributions to
renormalization constants, such that the Fock-space expansion of the physical state is practical. At this point the quasi-potential framework with { dressed particles}, namely with vacuum polarization effects, allows to introduce the valence component of the LF wave function, while keeping the full content of the Minkowski space BS formulation of the bound state, to the kernel of the effective interaction of in the square mass eigenvalue equation for the valence { component}. Any observable computed with the BS amplitude, can be systematically expanded  in a way to include the contribution beyond the valence. For bosonic theories, an essential step was the introduction of the dressed disconnect four-point LF Green's function given by Eq.~\eqref{2.11}, and for two fermions is written in Eq.~\eqref{3F1}, at the expense of introducing a Dirac projector on LF  spinors with running mass, that for the moment is introduced to allow the quasi-potential expansion of the BSE, and the definition of a valence { component}. On the practical side it was noted in~\cite{Moita:2022lfu} that in for the pion bound state with dressed quarks, the standard formula do derive the valence LF wave function from the BS amplitude as provided e.g. in ~\cite{dePaula:2020qna} does not refer to a explicit quark mass, although it has been derived starting with the LF quantization with constituent degrees of freedom. Therefore, the fermion self-energy introduced in \eqref{eq:dressproj} does not appear explicitly once the BS amplitude of the $0^-$  state of the pion is decomposed in the Dirac invariants~\cite{dePaula:2020qna}. Of course such a relation, can be further elaborated between the valence wave function and the one derived from the quasi-potential approach, that we leave open for a future work.

\section{Hierarchy equations for the valence LF Green’s functions}\label{sec:hierarchy}

The projection of the dynamics of the valence { component} onto the LF, through the associated two-body Green's functions, allows for an in-depth exploration of the Fock-space content that was implicit in the quasi-potential method and in the effective interaction \( w(K) \) introduced in the previous sections. In what follows, we will recall the discussions presented in Refs.~\cite{Frederico:2003zk,Frederico:2010zh}, where the example of the LF projection of the ladder BSE was explored to expose the coupling of the valence sector with higher Fock components, as previously studied for Hamiltonians considering off-diagonal matrix elements in the LF Fock space~\cite{Pauli:1998tf,Pauli:1999hi}. Here, the discussion  will now  be placed in the context of dressed particles, developed in sections~\ref{sec:global2boson} and \ref{sec:global2fermion}. The relevant point is that Fock space expansion is performed on the dressed states of each individual boson, namely particles 1 and 2. The example to be explored has the coupling of these dressed bosons to an intermediate bare one in ladder approximation. The reader has to be warned that we are not dealing with the full solution of the quantum field theory, but illustrating how the dressed constituents can come together to form the Fock-space decomposition with { dressed particles} .

The starting point is the disconnected LF resolvent operator for two dressed bosons written as
\begin{eqnarray}
g_0(K)=| G_0(K)|:=  \int dk^{\prime -}_1 dk^{ -}_1
\left\langle k_1^{\prime -}\right|
G_0(K)\left|k^-_1\right\rangle \
=i\widehat\Omega^{-1}g_0^{(2)}(K)\widehat\Omega^{-1} \ , \label{eq5}
\end{eqnarray}
where $\widehat{\Omega} = \sqrt{\widehat{k}_1^+ (K^+ - \widehat{k}_1^+)}$ is the phase-space operator. 
The role of $\widehat{\Omega}$ is to ensure invariance of the LF projection (``$|$'') under the kinematical LF subgroup of the Poincaré group.

Specializing to scalar particles, the two-body LF free Green's function is a particular case of the $N$-body version with the dressed bosons 1 and 2:
\begin{equation}
g_0^{(N)}(K) = \left[\prod_{j=1}^N \theta(\widehat{k}_j^+) \theta(K^+ - \widehat{k}_j^+)\right] \int_0^\infty ds_1^2\int_0^\infty ds_2^2 \,\frac{\rho(s_1^2)\,\rho(s_2^2)}{
K^- - \widehat{K}_0^{(N)-} + i\varepsilon}\ ,
\label{eq:nb-free}
\end{equation}
where $\widehat{K}_0^{(N)-} = \widehat{
k}_{1on}^{-}+\widehat{k}_{2on}^{-}+ (1-\delta_{N,2})\sum_{j=3}^N \widehat{k}_{j\text{on}}^-$ takes into account the dressing of the two particles, as one can check by comparing to Eq.~\eqref{2.11} for $N=2$, namely the disconnected global LF two-boson propagator presented in section~\ref{sec:global2boson}. 

In the two-particle sector, the interacting LF Green's function from Eq.~(\ref{LFRESOLV0}) can be recast as
\begin{equation}
g^{(2)}(K) = g_0^{(2)}(K) + g_0^{(2)}(K) \, \nu(K) \, g^{(2)}(K),
\end{equation}
where $g^{(2)} = -i \, \widehat{\Omega} \, g(K) \, \widehat{\Omega}$, and the effective interaction is
$\nu(K) = i \, \widehat{\Omega}^{-1} w(K) \widehat{\Omega}^{-1}\,.$
The leading contributions form the expansion of $w(K)$ to $\nu(K)$ are
 \begin{eqnarray}
&&
\nu^{(2)}(K)=i \left[\widehat\Omega g_0(K)\right]^{-1}|
G_0(K)V(K) G_0(K)| \left[g_0(K)\widehat\Omega\right]^{-1}\ ,
\label{eq14}
\\   &&
\nu^{(4)}(K) =i \left[\widehat\Omega g_0(K)\right]^{-1}| G_0(K)V(K)
G_0(K)V(K) G_0(K)|\left[g_0(K)\widehat\Omega\right]^{-1} \nonumber \\  &&
-i\left[\widehat\Omega g_0(K)\right]^{-1}| G_0(K)V(K)\widetilde
G_0(K)V(K) G_0(K)|\left[g_0(K)\widehat\Omega\right]^{-1} \,,
\label{eq16} \end{eqnarray}
where the subtraction in $\nu^{(4)}(K)$ guarantees two-body irreducibility.

To explore the LF Fock-sector content of $\nu(K)$, we use the example of a bosonic Yukawa-type interaction:
\[
\mathcal{L}_I^B = g_S \, \phi_1^\dagger \phi_1 \sigma + g_S \, \phi_2^\dagger \phi_2 \sigma,
\]
where $\phi_{1,2}$ and $\sigma$ are bosonic fields.
The interaction matrix element associated with emission/absorption of a $\sigma$ boson reads
\begin{eqnarray}\langle q k_\sigma |v|k\rangle= -2(2\pi)^3\delta^3 (\tilde
q+\tilde k_\sigma-\tilde k) \frac{g_S}{\sqrt{q^+k^+_\sigma k^+}}
\theta (k^+_\sigma)~ , \label{eq18} \end{eqnarray}
with LF momentum components $\tilde{q} = (q^+, \mathbf{q}_\perp)$. Note that only the kinematical momentum of the bosons 1 and 2 enters in computing the matrix element, and we are so far not dressing the interaction vertices.

Keeping terms up to second order in $v$, the effective interaction becomes
\begin{equation}
\nu(K) \approx \nu^{(2)}(K) + \nu^{(4)}(K) = v g_0^{(3)}(K) v + v g_0^{(3)}(K) v g_0^{(4)}(K) v g_0^{(3)}(K) v.
\label{eq:nu_expanded}
\end{equation}
where $v$ represents the basic interaction vertex.  The resolvents $g_0^{(3)}(K)$ and $g_0^{(4)}(K)$ are the three- and four-particle free global LF propagators, respectively.

Generalizing, one arrives at
$\nu(K) = v \, g^{(3)}(K) \, v,$
where $g^{(3)}(K)$ itself includes the virtual LF propagation in higher Fock states, leading to a hierarchy of coupled equations:
\begin{eqnarray}
&&g^{(2)}(K)=g^{(2)}_0(K) +g^{(2)}_0(K)v
g^{(3)}(K)vg^{(2)}(K)~ \ . \ . \ . \nonumber \\  &&
 g^{(N)}(K)=g^{(N)}_0(K) +g^{(N)}_0(K)v g^{(N+1)}(K)vg^{(N)}(K)~
 \ . \ . \ . \ \label{eq22}
\end{eqnarray}
The above set of coupled equations encode the Fock-space organization of the LF Quasi-Potential expansion. It is worthwhile to mention that the truncation of the effective interaction $\nu$ in the QP expansion is different from truncating  the coupled set of Eqs.~\eqref{eq22} in the Fock-space. This can be understood by restricting
the intermediate-state propagation up to four-particles, namely retaining up
to $g^{(4)}_0(K)$. Then,  one gets
\begin{equation}\label{eq:g2}
g^{(2)}(K)\simeq g^{(2)}_0(K) +g^{(2)}_0(K)v g^{(3)}(K)vg^{(2)}(K)\ ,
\end{equation}
with 
\begin{equation}\label{eq:g3}
g^{(3)}(K)\simeq g^{(3)}_0(K) +g^{(3)}_0(K)v
g^{(4)}_0(K)vg^{(3)}(K)\,,
\end{equation}
where one has intermediate virtual LF propagation of up to four particles due to the presence of  $g^{(3)}(K)$. On the other side, from Eq.~\eqref{eq:nu_expanded},
one has at the second order in the QP expansion only the term
$v g_0^{(3)}(K) v g_0^{(4)}(K) v g_0^{(3)}(K) v$ without solving  Eq.~\eqref{eq:g3} for $g^{(3)}(K)$, which includes the coupling up to four-particle states.

 We address the reader to Refs.~\cite{Frederico:2003zk,Frederico:2010zh}, where it was discussed the coupled set of equations for the LF Green's function for an interacting system composed by two undressed fermions. The detailed analysis of  the interesting fermionic problem with dressed constituents is left for a future study.
 
\section{LF projection of the Faddeev-Bethe-Salpeter equation { with dressed particles}}
\label{sect:FBSLF}

The methodology outlined in Sects.~\ref{QPred}, \ref{sec:QPMLFProj} and \ref{lfval} and \ref{lfval} has been extended to three-particle systems, as discussed in Refs.~\cite{Marinho:2007zz,Marinho:2008zza,Frederico:2010zh,Guimaraes:2014kor,Ydrefors:2021mky}. Three-body systems on the LF have been explored through zero-range models~\cite{Frederico:1992np,Carbonell:2002qs},  and employed in the description of the nucleon~\cite{deAraujo:1995mh,Ydrefors:2021mky}. Such model also has been applied to study heavy baryons~\cite{Suisso:2002jg}.
Worthwhile to mention that efforts has been taken to study the nucleon within BLFQ~\cite{Xu:2021wwj} and  now including besides $|3q\rangle$ also $|3q\,g\rangle$ and $|4q\bar q\rangle$ Fock states in the decomposition of the nucleon LF wave function~\cite{Xu:2024sjt}. Other approaches dealing with  Poincar\'e covariant Faddeev-Bethe-Salpeter equations  have been explored in the literature to study the structure of the nucleon and other baryons~\cite{Eichmann:2016yit,Segovia:2015hra,Eichmann:2018adq,Liu:2023reo,Chen:2023zhh} including recently use of contact interaction~\cite{Yu:2024qsd}.

The derivations presented in this section follows closely Ref.~\cite{Ydrefors:2021mky} with the novelty of including dressed particles, which are now introduced in the Faddeev-Bethe-Salpeter (FBS) equations and projected to the light-front. { The physical motivation to develop the FBS formalism and its projection to the light-front with dressed particles is justified by the desire to take into account the dressed quark propagator, where the self-energies are momentum dependent (see e.g. ~\cite{Oliveira:2018fkj}) once spin is included, bringing the physics of the spontaneous breaking of the chiral symmetry in the computation of the nucleon structure observables, like e.g., the multitude of parton distributions. }

The present discussion is based on the QP expansion  as has ben done for  two bosons systems presented in sect.~\ref{lfval} and here we sketch  the main steps in deriving the LF projected FBS equations for the valence { component} and the corresponding integral equation for the vertex function, obtained in leading order on the QP expansion, that corresponds to the one derived in originally in Ref.~\cite{Frederico:1992np} for the contact interaction.

We start with the three-boson  BS amplitude defined by
\begin{equation} \label{Phi0} \Psi_M(y_1,y_2,y_3;p)=\langle 0 \left| T(\varphi (y_1)\varphi (y_2)\varphi (y_3))\right| K\rangle \, ,\end{equation}
where $y_i$ is the space-time position of  particle $i$, $\varphi(y)$ is the bosonic field operator and $K$ the total momentum.

Following the two particle case, the valence LF wave function is derived from the BS amplitude as its projection onto the light-front: 
\begin{equation} \label{bs7b}
\psi_3 (\vec{k}_{1\perp},x_1;\vec{k}_{2\perp},x_2;\vec{k}_{3\perp},x_3)=(p^+)^2 \, (x_1 x_2 x_3)^\frac12
\chi_3 (\vec{k}_{1\perp},x_1;\vec{k}_{2\perp},x_2;\vec{k}_{3\perp},x_3)\, ,
\end{equation}
with
\begin{equation} \label{bs7c}
\chi_3 (\vec{k}_{1\perp},x_1;...)=
\int dk^{-}_{1}\, dk^{-}_{2}\, 
\Phi_M(k_1,k_2,k_3;p)\, ,
\end{equation}
where the double integration on $k^-$ eliminates the relative LF time between the three-particles.
The representation of the Minkowski space BS amplitude $\Psi_M$ in momentum space is $\Phi_M$. The auxiliary LF amplitude $\chi_3$ is used for the  convenience of our notation. The following operation represents the LF projection for a quantity $A$ defined in Minkowski space:
\begin{eqnarray}
|A:=\int dk_1^-dk_2^-\langle k_1^-k_2^-|A,\quad\text{and}\quad
A|:=\int dk_1^-dk_2^-A|k_1^-k_2^-\rangle,
\label{bardef3}
\end{eqnarray}
and note that the  matrix element of $A$ depends only on two independent momenta after the total momentum is factorized out. Using the "$|$" operation, one can write that:
\begin{equation}
|\chi_3\rangle=\big||\Phi_M\rangle=\big|G_0|\Gamma_M\rangle\quad \text{and} \quad |\Gamma_M\rangle= V  G_0 |\Gamma_M\rangle\ ,
\end{equation}
with  the last one being the BSE for the vertex function.

Explicitly the  three-particle free disconnected Green's function for dressed bosons, that comes as a generalization of the two-boson one, Eq.~\eqref{3},  is written as:
\begin{multline}\label{eq:g03b}
\langle k_1^-,k_2^- |G_0|k_1^{\prime -},k_2^{\prime -}
\rangle=\frac{-i}{(2\pi)^2}\frac{\delta(k_1^--k_1^{\prime -})\,\delta(k_2^--k_2^{\prime -})}
{\hat{k}_1^+\hat{k}_2^+(K^+-{\hat{k}_1}^+-\hat{k}_2^+)}\\ \times \left[\prod_{i=1}^3\int_0^\infty ds_i^2\,\rho(s_i)\right]~ \frac{1}
{(k_1^--\hat{k}_{1 on}^{s-})(k_2^--\hat{k}_{2
on}^{s-})(K^--k_1^--k_2^--\hat{k}_{3on}^{s-})}~,
\end{multline}
where the on-minus-shell momentum $ \hat{k}^{s-}_{ion}=(\hat{\vec k}^2_{i\perp}+s^2_i)/\hat{k}_i^+$.
Momentum conservation applies such that $\hat k^+_3=K^+-\hat{k}^+_1-\hat{k}^+_2$  and the analogous expression for the transverse components.  The LF projection of Eq.~\eqref{eq:g03b} is
    $g_0 = |G_0|,$
being the free light-front resolvent, explicitly written as:
\begin{multline}
g_0(\tilde{k}_1,\tilde{k}_2)
=\frac{i\theta(K^+-k_1^+-k_2^+)\theta(k_1^+)\theta(k_2^+)}{k_1^+k_2^+(K^+-{k_1}^+-k_2^+)} 
\\ \times \left[\prod_{i=1}^3\int_0^\infty ds_i^2\,\rho(s_i)\right]~ \,\frac{1}{K^--k_{1 on}^{s-}-k_{2 on}^{s-}-(K-k_1-k_2)_{on}^{s-}}
~,\label{propfl3}
\end{multline}
where $\tilde{k}_i\equiv\{ 
k^+_i,\vec k_{i\perp}\}$. The auxiliary Green's function is $ \widetilde{G}_0 = G_0|g^{-1}_0|G_0$ 
and with that the four-dimensional BSE for the vertex function is also a solution of:
\begin{equation}
  |\Gamma_M\rangle= W  \widetilde G_0 |\Gamma_M\rangle=W G_0|g^{-1}_0|G_0|\Gamma_M\rangle\, ,
\end{equation}
where the quasi-potential is 
$W = V + V \Delta_0 W$
with $ \Delta_0 = G_0 - \widetilde{G}_0\, .$

The LF auxiliary amplitude is written as:
\begin{equation}
  |\chi_3\rangle=\big|G_0   |\Gamma_M\rangle=\big |G_0 W G_0\big|g^{-1}_0|G_0\big|\Gamma_M\rangle\, ,
\end{equation}
where the LF interaction for the valence { component} is
   $ w = g_0^{-1}|G_0 W G_0| g_0^{-1}\,,$
leading to $|\chi_3\rangle=g_0\,w|G_0\big|\Gamma_M\rangle =
  g_0\,w|\chi_3\rangle \, ,$
and  the LF vertex function is a solution of
  $  |\Gamma_{LF}\rangle =w\, g_0|\Gamma_{LF}\rangle\, .$

In  what follows the Faddeev decomposition of the vertex function is introduced, for that the potential is built from the two-body ones as (see e.g.~\cite{Guimaraes:2014kor}):
\begin{equation}
\label{{Eq:V}}
    V = \sum_{i=1}^3 V_i,\quad\text{and}\quad V_i = V_{(2)i}S^{-1}_i
\end{equation}
where $S_i$ is the dressed propagator of the particle $i$ and $ V_{(2)i}$ is the interaction between the particles $j$ and $k$. The Faddeev decomposition of $W$ is:
\begin{equation}
\label{Eq:Faddeev_W}
    W = \sum_{i} W_i \quad\text{and}\quad w=\sum_{i}w_i,
\end{equation}
where
$ w_i = g_0^{-1}|G_0 W_i G_0| g_0^{-1}\, .$.  Then, the valence LF three-body wave function is decomposed as:
\begin{equation}
  |\chi_3\rangle =
  g_0\,\sum_i|v_i\rangle
  \quad\text{with}\quad |v_i\rangle=w_i|\chi_3\rangle\, ,
\end{equation}
where $|v_i\rangle$ are the Faddeev components of the vertex function, $|\Gamma_{LF}\rangle=\sum_i|v_i\rangle$.

The expansion of the Faddeev component of the QP interaction is:
\begin{equation}
\label{Eq:W_exp}
\begin{aligned}
    W_i &= V_i +V_i \Delta_0(V_i + V_j + V_k) \\
    & + V_i\Delta_0(V_i + V_j + V_k)\Delta_0(V_i + V_j + V_k) + \cdots 
    \end{aligned}
\end{equation}
and at the leading order (LO)  the effective potential for the valence { components} is:
\begin{equation}
\label{Eq:w_L1}
    w^{\text{LO}}_i = g_0^{-1}|G_0 V_i G_0| g_0^{-1} \, ,
\end{equation}
allowing to build the Faddeev equations for the components of the valence { component} vertex in LO in the QP expansion:
\begin{eqnarray}
    \label{Eq:v_eq_2}
  |v^{\text{LO}}_i \rangle&=&  t^{\text{LO}}_i g_0 (|v^{\text{LO}}_j\rangle + |v^{\text{LO}}_k\rangle)\, ,  
\end{eqnarray}
with the LF T-matrix satisfying a Lippman-Schwinger type integral equation:
\begin{equation}\label{Eq:tLO}
    t^{\text{LO}}_i = w^{\text{LO}}_i+w^{\text{LO}}_i g_0\,t^{\text{LO}}_i\,,
\end{equation}
where $g_0$ and the effective interaction are immersed in the three-body system, analogous what is found in the non relativistic quantum mechanics formulation.

Our formulation of the three-boson system now involves dressed particles, and considering the contact interaction, to illustrate the new equation, we introduce the
matrix element of the potential $V_i$ as:
\begin{equation}
\label{Eq:V_me}
    \langle k_j, k_k |V_i|k'_j, k'_k \rangle = \lambda (2\pi)^2 \delta(k_i - k'_i)(k_i^2 - m^2)\, .  
\end{equation}
and introducing this interaction in Eq.~\eqref{Eq:w_L1}, one should solve the LF T-matrix equation~\eqref{Eq:tLO}, where the  spectral densities of the dressed propagators have to be taken into account. We just sketch the expected general form:
\begin{equation}
\label{Eq:me}
    \Bigl\langle  \tilde k_j, \tilde k_k \Bigl|t^{\text{LO}}_i \Bigr|  \tilde k'_i, \tilde k'_j\Bigr\rangle =
    -i\mathcal{F}(\tilde k_i;K) k^+_i \delta(\tilde k_i - \tilde k'_i),
\end{equation}
where momentum conservation implies that $\tilde K=\tilde k_i+\tilde k_j+\tilde k_k=\tilde k'_i+\tilde k'_j+\tilde k'_k\, .$  The solution of Eq.~\eqref{Eq:tLO} for $\mathcal{F}(K,\tilde k_i)$ requires the knowledge of the spectral densities, which can be interesting by itself but beyond our presentation, and we leave that for a future exploration. However, we can continue and present the general form of the LF FBS for the valence { component} at LO in the QP expansion.

Notice that  $V_i\Delta_0 V_i = 0$ for the zero-range interaction  implying that $W_i = V_i$ at the leading order, therefore:
\begin{equation}
\label{Eq:v_eq_3}
    v^{\text{LO}}_i(\tilde k_i) = -2i \mathcal{F}(\tilde k_i;K) \int d\tilde k'_j k^+_i g_0(\tilde k'_i, \tilde k'_j)v^{\text{LO}}_j(\tilde k'_j)\, ,
\end{equation}
where the measure is
$d\tilde k\equiv\frac{dk^+d^2k_\perp}{2(2\pi)^3}$ and 
the factor of two comes from the symmetrization of the total vertex function with respect to the exchange of the bosons.  In the case of bare bosons, Eq.~\eqref{Eq:v_eq_3} simplifies and the Faddeev equation for the LF valence vertex has been written in~\cite{Frederico:1992np,Carbonell:2002qs} and more recently revisited in~\cite{Ydrefors:2020duk}. For the three-body continuum  the Minkowski space Faddeev equations for the $3\to 3$ scattering were derived in~\cite{Magalhaes:2011sh} and the LO projected equations in
\cite{Guimaraes:2014kor}, where these frameworks were applied to the study of
final state interaction in $D^{+} \to K^{-} \pi^{+} \pi^{+}$ considering the $K\pi$ isospin 1/2 and 3/2  channels. More recently, the continuum four-dimensional Minkowski space Faddeev equations were derived for the particular case to take into account the final state interaction in $B^+\to p\bar p\pi^+(K^+)$ decays~\cite{Bediaga:2024ipi}.

 \section{Minkowski Space BSE and Nakanishi Representation}\label{sect: MinkBSE}

The QP method applied to the Minkowski space BSE  has been reviewed in Sect.~\ref{sec:BSE-LFP} and \ref{sect:FBSLF} for the case of two and three-particles aiming to project it onto the light-front. The projection of the BSE to the LF inherently encompass the coupled dynamics of the valence { component} with an  infinite set of Fock components, as discussed in the cases of bound and scattering states, even when  the ladder approximation is adopted with bare and { dressed particles}  as has been elaborated in Sect.~\ref{sec:hierarchy}. In this perspective it is appealing to  solve the BSE directly in Minkowski space to access the rich Fock content of an interacting bound/scattering state. Worthwhile to mention that within such perspective, the hadron state in QCD  should be composed by  infinitely many LF Fock-components even at the initial hadronic scale (see e.g.~\cite{Brodsky:1997de}). 

However, the direct solution of the BSE in Minkowski space is a  challenging problem in quantum field theory. The complexities arise from the intricate analytic structure of Green’s functions and the involved amplitudes, which exhibit singularities on the real energy axis that usually hinders straightforward numerical treatments.  On the other side, these singularities has been overcome by resorting to formulation in Euclidean space that however limit the accessibility of physical observables such as timelike form factors, decay constants, and scattering amplitudes, beyond the perturbative expansion, that are crucial to explore the rich phenomenology of QCD, which requires to make use of non perturbative methods. 
Overcoming these issues in Minkowski space requires techniques that are able to deal with the singularities appearing in the kernel of the BSE.

There are different methods that can be used to solve the Minkowski space BSE using the Nakanishi integral  representation~\cite{Kusaka:1995vv} and further elaborated 
to solve ladder and cross-ladder approximations to the BSE for both scalar and fermionic theories in a series of works~\cite{Sauli:2001we,Carbonell:2006zz,Carbonell:2005nw,Carbonell:2010zw,Frederico:2011ws,Frederico:2013vga,dePaula:2016oct,Karmanov:2021imh,Jia:2023imt}; a method to directly solve the BSE in Minkowski space based on a proper treatment of the singularities appearing in the kernel and in the BS amplitude itself
\cite{Carbonell:2013kwa,Carbonell:2014dwa,Ydrefors:2020duk} and
by contour deformation~\cite{Eichmann:2021vnj}, a technique also applied to compute  three-point functions of the $\phi^3$
model~\cite{Huber:2022nzs}.

{ The goal of this section is two-fold: (i) generalize the method based on the Nakanishi integral representation to solve the BS equation with dressed particles, and (ii) show the formal equivalence between the eigenvalue equation for the valence component of the light-front wave function, Eq.~\eqref{lfbse}, developed in Sect.~\ref{lfval} and the integral equation derived using the NIR.}

 The NIR transforms the BSE into an equation for a weight function that captures the dynamics of the bound state while isolating the kinematic singularities into known analytical structures. In the two-body case, the BSE amplitude \( \Phi(k, K) \) for constituents with momenta \( k \) and total momentum \( K \), is represented as:
The BS amplitude for an $s-$wave state 
can be written in  terms of NIR as 
\begin{equation}\label{NIR-bs}
\Phi(k,K)=-{\rm i} \int_{-1}^{1} dz \int_{0}^{\infty} d \gamma \frac{g(\gamma,z;K^2)}
{[\gamma-k^2+ K \cdot k\,z-{\rm i}\epsilon]^n}, 
\end{equation}
where \( g(\gamma, z) \) is the Nakanishi weight function, and \( n\) is an integer depending on the problem at hand. This representation holds for scalar, fermionic, and vector systems and allows direct treatment in Minkowski space where the  \( g(\gamma, z) \)  is free of singularities in the real domain of its variables. It is  amenable to  be expanded in a suitable basis, such as Gegenbauer or Laguerre polynomials, and solving the linear equation for the expansion coefficients numerically, and in this way one can compute bound-state masses and amplitudes with high accuracy. This method preserves the full relativistic structure and avoids to deal with analytic extrapolations typically involved in Euclidean-space formulations.

The method introduced in ~\cite{Carbonell:2006zz} to solve the BSE allies the projection onto the light-front  after introducing the NIR representation of the BS amplitude. Such technique works in practice, and the reason for that is the LF projected BS amplitude starting from the Nakanishi representation, is a Stieltjes transform~\cite{Carbonell:2017kqa}, which is invertible once the kernel of the BSE can be written in a Nakanishi form~\cite{Frederico:2011ws}. This solves the issue of uniqueness of the solution for the Nakanishi weight function, as it has only be proven for perturbatively, and could be questionable to be used in non perturbative problems as the bound state. This observation  leads to the existence of the interacting reverse LF-projection operator, Eq.~\eqref{invop}, introduced through the QP expansion in Sect.~\ref{lfval} to obtain the full Minkowski space BS amplitude from the valence { component}.

To substantiate the above discussion  the general formalism to solve the homogeneous BSE in Minkowski space as introduced in Ref.~\cite{Carbonell:2006zz} and further explored in~\cite{Frederico:2011ws,Frederico:2013vga} is reviewed in what follows.
The BSE for the relativistic bound state with mass $M=\sqrt{K^2}$ and dressed particles is:
\begin{equation}\label{bse}
\Phi(k,K)= G_{0}(k,K) \int \frac{d^4 k'}{(2\pi)^4} {i} \,
V(k,k';K)\,\Phi(k',K) \, ,
\end{equation}
where ${\rm i} \,V(k,k';K)$  contains all two-body irreducible diagrams, and, as we have written in Eq.~\eqref{2a} for the two dressed spinless particles, the disconnected propagator is:
\begin{equation}\label{free-prop}
G_{0}(k,K)=i^2\int_0^\infty ds_1\int_0^\infty ds_2 \frac{\rho_1(s_1)}{\left[(K/2+k)^2 -s_1^2 +{\rm i}\epsilon\right]} \frac{\rho_2(s_2)}{\left[(K/2-k)^2 -s_2^2 +{\rm i}\epsilon\right]}
,\end{equation}
where the factors are written to make consistent the notation used in this section.

After substituting Eq.~(\ref{NIR-bs}) in (\ref{bse}) and integrating over $k^{-}$ on both sides,  one can
obtain the following  homogeneous integral equation for the Nakanishi weight function with $n=3$ in Eq.~\eqref{NIR-bs}:
\begin{equation}\label{nakie}
\int_0^{\infty}d\gamma'
\left\{
\frac{g( \gamma, z; K^2)}
{[ \gamma' + \gamma +\frac14 z^2 K^2]^{2} }-\int_{-1}^{1}dz' V^{LF}(z,z',\gamma, \gamma';K) \,g(\gamma',z';K^2)\right\}
=0
,\end{equation}
where
\begin{equation} \label{val1}
\int_0^{\infty}d\gamma'~\frac{g(\gamma',z;\kappa^2)}
{[\gamma'+\gamma - \frac14 z^2 K^2]^2}=K^+
\int {dk^- \over 2 \pi} \Phi(k,K)=
{\frac{ \psi(\xi,{\bf k}_\perp)} {\xi(1-\xi)}}, 
\end{equation}
with $\gamma=|{\bf k}_\perp|^2$, $\xi=(1 -z)/2$ and $\psi(\xi,{\bf k}_\perp)$ is the {\em valence component of the
light-front wave function}. The left-side of Eq.~\eqref{val1} is a Stieltjes transform (see Ref.~\cite{Carbonell:2017kqa}).

The Nakanishi kernel, $V^{LF}$,  is given in terms of the BSE kernel, by
\begin{small}
 \begin{equation}
V^{LF}(z,z',\gamma, \gamma';K)\equiv {\rm i} K^{+} \int_{-\infty}^{\infty} \frac{dk^{-}}{2 \pi} G_{0}(k,K) \int \frac{d^4 k'}{(2 \pi)^4} 
\frac{{\rm i}V(k,k';K)}{[k'^2 + K\cdot k'\,z'-\gamma' + {\rm i}\epsilon ]^3}\,
.\end{equation}
\end{small}
 In the case of identical bosons
the Nakanishi weight function has the property of 
$g(\gamma,z;\kappa^2)=g(\gamma,-z;\kappa^2)$ as requested by the bosonic nature of these constituents.  
In this case, the $z^2$ dependence of $g(\gamma,z;\kappa^2)$ entails the symmetry of
 the valence wave function, namely $\psi(\xi,{\bf k}_\perp)=\psi(1-\xi,{\bf k}_\perp)$. We point out that in the situation that the BSE interaction kernel, $V(k,k';K)$, is written as two-body irreducible diagrams, it is assured that the Nakanishi kernel will be written as Stieltjes transform (see ~\cite{Frederico:2011ws,Frederico:2013vga}), and therefore  Eq.~\eqref{nakie} is translated to:
\begin{equation}\label{nakie1}
g( \gamma, z; K^2)
=\int_0^{\infty}d\gamma'
\int_{-1}^{1}dz'~ \widetilde V^{LF}(z,z',\gamma, \gamma';K) \,g(\gamma',z';K^2)
\ ,\end{equation}
where the kernel $ \widetilde V^{LF}(z,z',\gamma, \gamma';K)$. In Ref.~\cite{Nakanishi:1971} it was shown that any sum of perturbative diagrams, including loops, can be written as an integral  in a Nakanishi integral form, based on the recurrent application of the  Feynman parametrization technique to have single denominator, and therefore after integration over $k^-$, the whole kernel of the BS equation becomes a Stieltjes transform. In Ref.~\cite{Karmanov:2021imh}, by means of the inverse Stieltjes transform, it was derived a practical form of the kernel of Eq.~\eqref{nakie1} in ladder plus cross-ladder approximation, that makes calculation of the  Minkowski Bethe-Salpeter amplitude no more difficult than the Euclidean one. Furthermore, the excited states can be computed by means of  Eq.~\eqref{nakie}
(see \cite{Gutierrez:2016ixt}). 
Another aspect of Eq.~\eqref{nakie} is that it can be in principle related to the homogeneous  equation for the valence { component} within the QP approach once the Stieltjes transform is inverted to write down the Nakanishi weight in terms of the valence wave function and formally Eq.~\eqref{nakie} can be written  
as
\begin{equation}\label{nakie_1}
\psi(\xi,\mathbf{k}_\perp)=\xi(1-\xi)\int_0^{\infty}d\gamma'\int_{-1}^{1}dz' V^{LF}(z,z',\gamma, \gamma';K) \,\widehat L^{-1}\big[\psi\big](\gamma',z')
\,,\end{equation}
where $\widehat L^{-1}$  represents the inverse Stieltjes transform~\cite{Schwarz:2004mv}
which could give an insight on the full kernel of Eq.~\eqref{lfbse}, namely, the quasi-potential bound state equation. { We emphasize that the very existence of the mathematically exact inversion of the Stieltjes transform says that the equations derived by using the projection of the BSE onto the light-front are well defined (see~\cite{Carbonell:2017kqa}).}

The application of NIR to fermionic bound-state problems, using light-front projection via
$k^-$ integration on both sides of the BSE~\eqref{bse} leading to Eq.\eqref{nakie}, is hindered by end-point singularities. These singularities can be effectively handled using the Yan integral~\cite{Yan:1973qg}, as demonstrated in the Feynman gauge in~\cite{dePaula:2016oct,dePaula:2017ikc}, where also a form-factor was introduced in the fermion-vector vertex. The dressing of the fermions does not introduce additional singularities, as detailed in~\cite{Castro:2023bij}. Furthermore, for fermion-scalar systems, end-point singularities are absent in the case of vector exchange within the ladder approximation~\cite{AlvarengaNogueira:2019zcs,Noronha:2023hjg}. We point out that the use of the inverse Stieltjes transform to workout the two-fermion BSE for the NIR, in analogy to Eq.~\eqref{nakie1} is not yet developed.  

\section{Applications to hadron structure}\label{sec:hadron}

The relevance of NIR to QCD lies in its ability to accommodate dynamically dressed fermionic propagators \cite{Castro:2023bij,Moita:2022lfu} and vertices, such as those obtained from Dyson-Schwinger equations (DSEs) \cite{Sauli:2002tk,Mezrag:2020iuo,Duarte:2022yur} and functional renormalization group (FRG) methods (see e.g. \cite{Alkofer:2018guy,Dupuis:2020fhh}). These non-perturbative inputs are critical for describing phenomena like dynamical chiral symmetry breaking (DCSB), confinement, and the running of the strong coupling at low-energy scales.  A particularly important point is that solutions in Minkowski space give access to { some} observables that are inaccessible in Euclidean space, such as the electromagnetic form factors in the timelike region.  The NIR also permits a direct calculation of light-front observables, as the projection onto the light-front hypersurface can be performed analytically due to the known structure of the representation \cite{dePaula:2016oct}. We should also mention the study of the  dynamical color-suppression of non-planar diagrams in charged two bound states reflected through the valence parton distribution  and electromagnetic form factor~\cite{AlvarengaNogueira:2017cpt}. 

Furthermore, the analytic continuation to the complex plane  of NIR is straightforward. This characteristic enables access to analytic properties of physical amplitudes, such as dispersion relations, branch cuts and eventual poles in the complex plane.  It also allows the derivation of generalized parton distributions (GPDs) and transition form factors in different kinematical regions,  that bridge the gap between inclusive and exclusive processes (for a recent review see~\cite{Mezrag:2023nkp}).
Beyond mesons, the NIR and Minkowski BSE has been extended to three-body systems to model the nucleon \cite{Ydrefors:2022bhq,Noronha:2023hjg}, allowing for the exploration of  form factors, including the gravitational ones~\cite{Burkert:2023wzr}, generalized parton distributions, and access the structure of exotic multiquark states using diquark degrees of freedom~\cite{Barabanov:2020jvn}, as for example building a dynamical model of two-diquarks and a quark to describe   pentaquarks (for a review on the experimental information see ~\cite{PDG2024}).

The Fock-space content of the BSE solutions reviewed in sections~\ref{sec:BSE-LFP}, \ref{sec:hierarchy} and for three constituents in sect.~\ref{sect:FBSLF} with  { dressed particles}, when associated  with the NIR, opens the opportunity to explore the light hadrons taking into account dynamical chiral symmetry breaking relevant for their acquired mass~\cite{Roberts:2016vyn}. The hadron state will inherently contain an infinite number of LF Fock-components with { dressed particles}, while its  valence component can be computed without truncation from the Mikowski BS amplitude. In principle, it provides a direct tool to connect  non-perturbative QCD inspired dynamical models to hadronic observables.  These include form factors, distribution amplitudes, generalized parton distributions, and structure functions — quantities that are essential for decoding the internal structure of mesons and baryons. 

Elastic electromagnetic form factors are among the most widely studied hadronic observables and provide direct access to charge and current distributions in hadrons. It has been  explored within the light-front quantization, where the form factors are typically extracted from the plus component of the electromagnetic current \( J^+ \), which receives dominant contributions~\cite{Brodsky:1997de} and normally minimizes  in the Drell-Yan-West frame the effect of end-point singularities (see e.g.~\cite{deMelo:1997hh,deMelo:1999gn,deMelo:2003cb,deMelo:2012hj,deMelo:2016lwr}). In this frame, the initial and final hadron momenta differ only in their transverse components. It is worth mentioning that with NIR the contribution of the higher Fock states to the form factors are implicitly accounted. Furthermore, at large momentum transfers, it is expected by counting rules that the valence contribution to the form factors should be dominant~\cite{Lepage:1980fj}, while at low momentum it represents only a fraction of the normalization of the LF wave function. A first  study of the pion electromagnetic form factor computed with a Minkowski space BS amplitude~\cite{Ydrefors:2021dwa} obtained in ladder approximation, with dressed gluon exchange, dressed quark-gluon vertex and constituent quarks, showed a valence probability of about 70\% and the dominance of the valence contribution for square momentum transfers above 70~(GeV/c)$^2$ .
Noteworthy to mention that the dressed quark gluon vertex was largely explored in Euclidean space~(see ~\cite{Oliveira:2020yac, Oliveira:2018ukh,Oliveira:2018fkj}), and it is a crucial dynamical part to be accounted in the BSE to describe hadronic structure, as it enhances the infrared (IR) interaction region as  expected in QCD.  { One possibility to consider the IR strengthening of the kernel of the Minkowski space BSE  is to implement a quark-gluon vertex, where the form factors have an integral representation, such that in the space-like region they qualitatively reproduce what is found in \cite{Oliveira:2020yac} where constraints from lattice QCD calculations were used to obtain information on the longitudinal components of the quark-gluon vertex function. Qualitatively such physics was already included in the kernel of the pion Minkowski space BSE in Ref.~\cite{dePaula:2020qna}, where on the side of a massive gluon propagator, it was considered a form factor for the $\gamma^\mu$ component of the quark-gluon vertex, which enhanced the kernel in the IR in comparison to the UV momentum region. }

The Fock space LF wave function encodes all the information about the momentum distribution of partonic constituents in a hadron. Indeed, descriptions of the pion state in the LF Fock space has been developed within BLFQ with $|q\bar q\rangle$ and $|q\bar q\,g\rangle$ components having probabilities of 49.2\% and 50.8\%, respectively~\cite{Zhu:2023lst}. Also in a phenomenological approach it was parametrized the
$|q\bar q\rangle$, $|q\bar q\,g\rangle$, $|q\bar q\,q\bar q\rangle$ and   $|q\bar q\,gg\rangle$ components of the pion LF wave function from the parton distribution functions and the
electromagnetic form factor~\cite{Pasquini:2023aaf}. For the nucleon state the first steps on its  Fock decomposition were carried out within BLFQ, including the valence sector $|qq q\rangle$, and $|qqq\,g\rangle$ (see~\cite{Lin:2024ijo} and references therein) computed with constituent quarks and gluons. Despite the tremendous advances in the description of the nucleon in Euclidean space has been undertaken with continuous Dyson-Schwinger and Bethe-Salpeter methods~(see e.g.\cite{Roberts:2023lap,Mezrag:2023nkp}), and also with lattice QCD calculations~\cite{LatticePartonCollaborationLPC:2022myp,LatticeParton:2023xdl,FLAG:2024oxs}, still the decomposition of the nucleon state in the LF Fock-space is in its infancy and approaches handling it is worthwhile to be pursued in view of building its 3D image on the null-plane~(see e.g. \cite{AbdulKhalek:2021gbh}).

On the other hand, the relation between the BLFQ  and the Minkowski BSE approaches has been so far only explored through the comparison of the pion PDFs with constituent degrees of freedom~\cite{Lan:2024ais}.  Qualitatively the contribution of the BLFQ $|q\bar q\,g\rangle$ state and the beyond the valence contribution brought by the Minkowski BS amplitude~\cite{dePaula:2020qna,dePaula:2022pcb,dePaula:2023ver} to the quark PDF is qualitatively similar exhibiting a maximum for $0<x<1/2$, indicating that the longitudinal momentum of the quark is shared with  partons present in higher Fock states at the physical pion scale. In principle, the LF projection QP toolkit, may allow to reveal the higher Fock-components buried in the Minkowski space BS approach, which will pave the way to bridge with frameworks that explicitly deal with a truncated Fock-space description of the hadron state.

The necessity to build the desired bridge between BS and the LF Fock-space representation of hadrons  is even strengthened with the  advent of the new facilities for Electron-Ion colliders that will usher the field in a new era of precision hadron tomography~\cite{AbdulKhalek:2021gbh}. 
The formalism developed here is ideally suited to provide the non-perturbative theoretical foundation for interpreting  measurements of GPDs, TMDs, and polarized structure functions. Accurate theoretical modeling of these observables using light-front projected BSEs will be useful for extracting insights about orbital angular momentum, spin correlations, and three-dimensional hadronic imaging \cite{AbdulKhalek:2021gbh, Accardi:2012qut}.

{ An outline to implement DCSB in Minkowski space would start with the formulation of both the quark self-energy DSEs and the BSE for the Goldstone modes, with consistent kernels, which, in addition, should be written through NIR. This would allow the solution of these equations to have an integral representation. From that, DSEs and BSEs will be re-expressed as integral equations for the spectral densities and for the NIR weight functions of the BS amplitudes, respectively. In connection to light-front dynamics, for example, the valence component of the pion light-front wave function will be obtained by the projection of the BS amplitude onto the light-front, and the physics of the dressed quark propagators will be implicitly included (see e.g. \cite{Castro:2023bij} ). 
 The quantitative exploration of DCSB with the solution of the regulated rainbow ladder DSE at the QCD scale using integral representation in Ref.~\cite{Duarte:2022yur}, in principle, can be extended to formulate the kernel of the BS equation for the pion, allowing its solution in Minkowski space, which can be explored with NIR in future studies. It is expected that the far-reaching consequences of DCSB~\cite{Roberts:2016vyn,Roberts:2021nhw} on the properties of the pion and kaon will be incorporated in the Minkowski space framework.

 Furthermore, the axial Ward-Takahashi identity expressing the chiral symmetry and how it is broken in QCD (see e.g. Eq.~(2.24) in Ref.~\cite{Roberts:2021nhw}), should be valid in Minkowski space. For that, the DSE for the quark two-point function, and the inhomogeneous BSEs for the axial-vector and the pseudo-scalar vertices should be computed with consistent kernels.  These  BSEs can be, in principle,  projected onto the LF using the QP method developed in sections \ref{sec:BSE-LFP} and \ref{sec:hierarchy}, and the truncation of QP expansion is expected to violate the axial Ward-Takahashi identity to some extend, as its known that full covariance which is not the case when the BSEs are solved fully in the four-dimensional Minkowski space, applying for example the NIR but to the inhomogeneous case (see e.g. ~\cite{Frederico:2011ws}), while in section \ref{sect: MinkBSE} it was briefly explained the method for the homogeneous case. 
 }

\section{Conclusions and Outlook}\label{sec:conclusion}

We review the framework for studying  bound-state dynamics in Minkowski space, emphasizing the solution of the BSE by both light-front projection via quasi-potential approach and via the Nakanishi integral representation. The NIR approach maintains full Lorentz covariance and allows direct computation of bound-state amplitudes, the valence light-front wave functions, and hadronic observables in the physical space.

The light-front projection of the BSE yields the valence wave functions that naturally match the partonic description of hadrons. In addition, the BS amplitude enables direct connections with experimental observables including electromagnetic and gravitational form factors, parton distribution amplitudes (PDAs), generalized parton distributions (GPDs), and structure functions.  Applications include the electromagnetic form factors of the pion \cite{Ydrefors:2021dwa}, momentum distributions \cite{dePaula:2020qna}, unpolarized TMDs \cite{dePaula:2023ver}, PDFs \cite{dePaula:2022pcb},  and proton quark distributions \cite{Ydrefors:2022bhq}. In principle, the covariant yet computationally tractable nature of this framework allows for a realistic modeling of mesons and baryons using dressed quark propagators and vertex functions derived from DSEs and constrained by lattice QCD inputs.

Our framework is particularly timely in light of ongoing and upcoming experimental programs. The future Electron-Ion Colliders, Jefferson Lab 12 GeV upgrade, and COMPASS++/AMBER at CERN will deliver unprecedented precision in mapping the internal structure of hadrons. The methods reviewed here are well-positioned to provide the theoretical tools required for interpreting these data. In particular, the extraction of 3D images of hadrons through GPDs and TMDs, the determination of gravitational form factors and pressure distributions, are areas where the reviewed non-perturbative techniques  may be useful.

Looking forward, several promising directions remain open for further development:
\begin{enumerate}[(a)]

    \item Extension to Three-Body Systems: while the  techniques focuses on two-body bound states, the extension to the Minkowski space BSE  to three-body systems such as the nucleon is an exciting avenue. Techniques for handling Faddeev-type equations on the light-front are actively being developed and can benefit from the NIR approach.

\item Systematic Track of Higher Fock States: the current Minkowski space technique can be improved to explore the higher Fock components within the BS approach. This will allow to separate the contributions of sea quarks and gluon degrees of freedom at the hadron scale, which is especially interesting in our understanding of the hadron properties with {dressed particles} .

\item Computational Advancements: implementing more sophisticated  methods  and leveraging machine learning techniques for solving the BSE can lead to substantial progress in numerical efficiency and accuracy of the Minkowski space solutions with impact on the different structure observables.

\end{enumerate}

In summary, the solution of the Minkowski-space Bethe-Salpeter approach,  using the Nakanishi integral representation and the projection onto the light-front, offers a powerful bridge between fundamental QCD dynamics and experimentally accessible hadronic observables.  
It enables direct access to physical observables and can incorporate non-perturbative dressing effects. When combined with the light-front projection, the Minkowski BS framework becomes a powerful tool for describing hadronic structure in a way that is both covariant and phenomenologically rich. The continued development of this approach faces the challenge to tackle  dynamical mass generation and confinement rooted in QCD, giving in practice a tool to pave the way to bridge the Fock-space representation of the eigenstates of light-front Hamiltonian and Minkowski space Bethe-Salpeter approach.

\section*{Acknowledgements}
W. d. P. acknowledges the partial support of the National Council for Scientific and Technological Development (CNPq) (Grants No. 313030/2021-9 and 401565/2023-8). T. F. thanks the partial support from CNPq (Grant No. 306834/2022-7). We thank the S\~ao Paulo Research Foundation (FAPESP) (Grants 2023/13749-1 and 2024/17816-8), and the  INCT-FNA project 408419/2024-5.


\begin{thebibliography}{99}

\bibitem{Briceno:2017max}
R.~A. Brice\~no, J.~J. Dudek, and R.~D. Young.
\newblock {Scattering processes and resonances from lattice QCD}.
\newblock {\em Rev. Mod. Phys.}, 90(2):025001, 2018.

\bibitem{Ji:2020ect}
X. Ji, Y.-S. Liu, Y. Liu, J.-H. Zhang, and Y. Zhao.
\newblock {Large-momentum effective theory}.
\newblock {\em Rev. Mod. Phys.}, 93(3):035005, 2021.

\bibitem{Carbonell:2010zw}
J.~Carbonell and V.~A. Karmanov.
\newblock {Solving Bethe-Salpeter equation for two fermions in Minkowski
  space}.
\newblock {\em Eur. Phys. J. A}, 46:387--397, 2010.

\bibitem{Salpeter:1951sz}
E.~E. Salpeter and H.~A. Bethe.
\newblock A relativistic equation for bound-state problems.
\newblock {\em Phys. Rev.}, 84:1232, 1951.

\bibitem{Nakanishi:1971}
N. Nakanishi.
\newblock {\em Graph Theory and Feynman Integrals}.
\newblock Gordon and Breach, New York, 1971.

\bibitem{Nakanishi:1969ph}
N.~Nakanishi.
\newblock A general survey of the theory of the bethe-salpeter equation.
\newblock {\em Prog. Theor. Phys. Suppl.}, 43:1, 1969.

\bibitem{Kusaka:1995vv}
K.~Kusaka and A.~G. Williams.
\newblock {Solving the Bethe-Salpeter equation for scalar theories in Minkowski
  space}.
\newblock {\em Phys. Rev. D}, 51:7026--7039, 1995.

\bibitem{Carbonell:2006zz}
J.~Carbonell and V.~A. Karmanov.
\newblock {Solving Bethe-Salpeter equation in Minkowski space}.
\newblock {\em Eur. Phys. J. A}, 27:1--9, 2006.

\bibitem{Frederico:2011ws}
T. Frederico, G. Salm\`e, and M. Viviani.
\newblock {Two-body scattering states in Minkowski space and the Nakanishi
  integral representation onto the null plane}.
\newblock {\em Phys. Rev. D}, 85:036009, 2012.

\bibitem{Frederico:2013vga}
T. Frederico, G. Salm\`e, and M. Viviani.
\newblock {Quantitative studies of the homogeneous Bethe-Salpeter Equation in
  Minkowski space}.
\newblock {\em Phys. Rev. D}, 89:016010, 2014.

\bibitem{Salme:2017oge}
G. Salm\`e, W. de~Paula, T. Frederico, and M. Viviani.
\newblock {Two-Fermion Bethe-Salpeter Equation in Minkowski Space: The
  Nakanishi Way}.
\newblock {\em Few Body Syst.}, 58(3):118, 2017.

\bibitem{Jia:2023imt}
S. Jia.
\newblock {Direct solution of Minkowski-space Bethe-Salpeter equation in the
  massive Wick-Cutkosky model}.
\newblock {\em Phys. Rev. D}, 109(3):036020, 2024.

\bibitem{Pimentel:2016cpj}
R.~Pimentel and W.~de~Paula.
\newblock {Excited States of the Wick-Cutkosky Model with the
  Nakanishi Representation in the Light-Front Framework}.
\newblock {\em Few Body Syst.}, 57(7):491--496, 2016.

\bibitem{Sales:1999ec}
J.~H.~O. Sales, T.~Frederico, B.~V. Carlson, and P.~U. Sauer.
\newblock {Light front Bethe-Salpeter equation}.
\newblock {\em Phys. Rev. C}, 61:044003, 2000.

\bibitem{Dirac:1949cp}
P.~A.~M. Dirac.
\newblock {Forms of Relativistic Dynamics}.
\newblock {\em Rev. Mod. Phys.}, 21:392--399, 1949.

\bibitem{Perry:1990mz}
R.~J. Perry, A. Harindranath, and K.~G. Wilson.
\newblock {Light front Tamm-Dancoff field theory}.
\newblock {\em Phys. Rev. Lett.}, 65:2959--2962, 1990.

\bibitem{Brodsky:1997de}
H.-C.~Pauli S.~J.~Brodsky and S.~S. Pinsky.
\newblock Quantum chromodynamics and other field theories on the light cone.
\newblock {\em Phys. Rept.}, 301:299, 1998.

\bibitem{Bakker:2013cea}
B.~L.~G. Bakker et~al.
\newblock {Light-Front Quantum Chromodynamics: A framework for the analysis of
  hadron physics}.
\newblock {\em Nucl. Phys. B Proc. Suppl.}, 251-252:165--174, 2014.

\bibitem{Mezrag:2014jka}
C.~Mezrag, L.~Chang, H.~Moutarde, C.~D. Roberts, J.~Rodr\'\i{}guez-Quintero,
  F.~Sabati\'e, and S.~M. Schmidt.
\newblock {Sketching the pion's valence-quark generalised parton distribution}.
\newblock {\em Phys. Lett. B}, 741:190--196, 2015.

\bibitem{dePaula:2016oct}
W.~de~Paula, T.~Frederico, G.~Salm\`e, and M.~Viviani.
\newblock {Advances in solving the two-fermion homogeneous Bethe-Salpeter
  equation in Minkowski space}.
\newblock {\em Phys. Rev. D}, 94(7):071901, 2016.

\bibitem{Accardi:2012qut}
A.~Accardi et~al.
\newblock {Electron Ion Collider: The Next QCD Frontier}: {Understanding the
  glue that binds us all}.
\newblock {\em Eur. Phys. J. A}, 52(9):268, 2016.

\bibitem{AbdulKhalek:2021gbh}
R.~Abdul~Khalek et~al.
\newblock {Science Requirements and Detector Concepts for the Electron-Ion
  Collider}: {EIC Yellow Report}.
\newblock {\em Nucl. Phys. A}, 1026:122447, 2022.

\bibitem{Accardi:2023chb}
A.~Accardi et~al.
\newblock {Strong interaction physics at the luminosity frontier with 22 GeV
  electrons at Jefferson Lab}.
\newblock {\em Eur. Phys. J. A}, 60(9):173, 2024.

\bibitem{Dudek:2012vr}
J. Dudek et~al.
\newblock {Physics Opportunities with the 12 GeV Upgrade at Jefferson Lab}.
\newblock {\em Eur. Phys. J. A}, 48:187, 2012.

\bibitem{Burkert:2023wzr}
V.~D. Burkert, L.~Elouadrhiri, F.~X. Girod, C.~Lorc\'e, P.~Schweitzer, and
  P.~E. Shanahan.
\newblock {Colloquium: Gravitational form factors of the proton}.
\newblock {\em Rev. Mod. Phys.}, 95(4):041002, 2023.

\bibitem{Haag:1955ev}
R.~Haag.
\newblock {On quantum field theories}.
\newblock {\em Kong. Dan. Vid. Sel. Mat. Fys. Med.}, 29N12:1--37, 1955.

\bibitem{Polyzou:2021qpr}
W. Polyzou.
\newblock {Relation between instant and light-front formulations of quantum
  field theory}.
\newblock {\em Phys. Rev. D}, 103(10):105017, 2021.

\bibitem{Brodsky:2000ii}
S.~J. Brodsky, D.~S. Hwang, B.-Q. Ma, and I. Schmidt.
\newblock {Light cone representation of the spin and orbital angular momentum
  of relativistic composite systems}.
\newblock {\em Nucl. Phys. B}, 593:311--335, 2001.

\bibitem{Sales:2001gk}
J.~H.~O. Sales, T.~Frederico, B.~V. Carlson, and P.~U. Sauer.
\newblock {Renormalization of the ladder light front Bethe-Salpeter equation in
  the Yukawa model}.
\newblock {\em Phys. Rev. C}, 63:064003, 2001.

\bibitem{dePaula:2017ikc}
W.~de~Paula, T. Frederico, G. Salm\`e, M. Viviani, and R.
  Pimentel.
\newblock {Fermionic bound states in Minkowski-space: Light-cone singularities
  and structure}.
\newblock {\em Eur. Phys. J. C}, 77(11):764, 2017.

\bibitem{Carbonell:2017kqa}
J.~Carbonell, T.~Frederico, and V.~A. Karmanov.
\newblock {Bound state equation for the Nakanishi weight function}.
\newblock {\em Phys. Lett. B}, 769:418--423, 2017.

\bibitem{Diehl:2003ny}
M.~Diehl.
\newblock {Generalized parton distributions}.
\newblock {\em Phys. Rept.}, 388:41--277, 2003.

\bibitem{Belitsky:2005qn}
A.~V. Belitsky and A.~V. Radyushkin.
\newblock {Unraveling hadron structure with generalized parton distributions}.
\newblock {\em Phys. Rept.}, 418:1--387, 2005.

\bibitem{Vary:2009gt}
J.~P.~Vary et~al.
\newblock Hamiltonian light-front field theory in a basis function approach.
\newblock {\em Phys. Rev. C}, 81:035205, 2010.

\bibitem{Li:2015zda}
Y. Li, P. Maris, X. Zhao, and J.~P. Vary.
\newblock {Heavy Quarkonium in a Holographic Basis}.
\newblock {\em Phys. Lett. B}, 758:118--124, 2016.

\bibitem{Binosi:2014aea}
D. Binosi, L. Chang, J. Papavassiliou, and C.~D. Roberts.
\newblock {Bridging a gap between continuum-QCD and ab initio predictions of
  hadron observables}.
\newblock {\em Phys. Lett. B}, 742:183--188, 2015.

\bibitem{Roberts:2016vyn}
C.~D. Roberts.
\newblock {Perspective on the origin of hadron masses}.
\newblock {\em Few Body Syst.}, 58(1):5, 2017.

\bibitem{Jia:2018ary}
S.~Jia and J.~P.~Vary. Basis light front quantization for the charged light mesons with color singlet Nambu-Jona-Lasinio interactions.
{\em Phys. Rev. C,} 99:  035206, 2019. 

\bibitem{Tsujimaru:1997jt}
S. Tsujimaru and K. Yamawaki.
\newblock {Zero mode and symmetry breaking on the light front}.
\newblock {\em Phys. Rev. D}, 57:4942--4964, 1998.

\bibitem{Nakanishi:1976vf}
N. Nakanishi and K. Yamawaki.
\newblock {A Consistent Formulation of the Null-Plane Quantum Field Theory}.
\newblock {\em Nucl. Phys. B}, 122:15--28, 1977.

\bibitem{Eichmann:2015nra}
G. Eichmann, C.~S. Fischer, and W. Heupel.
\newblock {Four-point functions and the permutation group S4}.
\newblock {\em Phys. Rev. D}, 92(5):056006, 2015.

\bibitem{Sanchis-Alepuz:2015qra}
H. Sanchis-Alepuz and R. Williams.
\newblock {Probing the quark\textendash{}gluon interaction with hadrons}.
\newblock {\em Phys. Lett. B}, 749:592--596, 2015.

\bibitem{Lebed:2016hpi}
R.~E.~Mitchell R.~F.~Lebed and E.~S. Swanson.
\newblock Heavy-quark qcd exotica.
\newblock {\em Prog. Part. Nucl. Phys.}, 93:143, 2017.

\bibitem{dePaula:2020qna}
W.~de~Paula, E.~Ydrefors, J.~H. Alvarenga~Nogueira, T.~Frederico, and
  G.~Salm\`e.
\newblock {Observing the Minkowskian dynamics of the pion on the null-plane}.
\newblock {\em Phys. Rev. D}, 103(1):014002, 2021.

\bibitem{dePaula:2023ver}
W.~de~Paula, T.~Frederico, and G.~Salm\`e.
\newblock {Unpolarized transverse-momentum dependent distribution functions of
  a quark in a pion with Minkowskian dynamics}.
\newblock {\em Eur. Phys. J. C}, 83(10):985, 2023.

\bibitem{dePaula:2022pcb}
W.~de~Paula, E.~Ydrefors, J.~H. Nogueira~Alvarenga, T.~Frederico, and
  G.~Salm\`e.
\newblock {Parton distribution function in a pion with Minkowskian dynamics}.
\newblock {\em Phys. Rev. D}, 105(7):L071505, 2022.

\bibitem{Ydrefors:2021dwa}
E. Ydrefors, W. de~Paula, J. H. Alvarenga Nogueira, T.
  Frederico, and G. Salm\'e.
\newblock {Pion electromagnetic form factor with Minkowskian dynamics}.
\newblock {\em Phys. Lett. B}, 820:136494, 2021.

\bibitem{Lan:2024ais}
J. Lan, C. Mondal, X. Zhao, T. Frederico, and J.~P. Vary.
\newblock Gluonic contributions to the pion parton distribution functions.
\newblock {\em Phys. Rev. D} 111: L111903, 2025.

\bibitem{Gross:1993zj}
F.~Gross.
\newblock {\em {Relativistic quantum mechanics and field theory}}.
\newblock 1993.

\bibitem{Yan:1973qg}
T.-M. Yan.
\newblock {Quantum field theories in the infinite momentum frame. 4. Scattering
  matrix of vector and Dirac fields and perturbation theory}.
\newblock {\em Phys. Rev. D}, 7:1780--1800, 1973.

\bibitem{Aguilar:2019teb}
A.~C. Aguilar et~al.
\newblock {Pion and Kaon Structure at the Electron-Ion Collider}.
\newblock {\em Eur. Phys. J. A}, 55(10):190, 2019.

\bibitem{Roberts:2021nhw}
C.~D. Roberts, D.~G. Richards, T. Horn, and L. Chang.
\newblock {Insights into the emergence of mass from studies of pion and kaon
  structure}.
\newblock {\em Prog. Part. Nucl. Phys.}, 120:103883, 2021.

\bibitem{AlvarengaNogueira:2019zcs}
J.~H. Alvarenga~Nogueira, D.~Colasante, V.~Gherardi, T.~Frederico, E.~Pace, and
  G.~Salm\`e.
\newblock {Solving the Bethe-Salpeter Equation in Minkowski Space for a
  Fermion-Scalar system}.
\newblock {\em Phys. Rev. D}, 100(1):016021, 2019.

\bibitem{wolja}
R.~M. Woloshyn and A.~D. Jackson.
\newblock Comparison of three-dimensional relativistic scattering equations.
\newblock {\em Nuclear Physics B}, 64:269--284, 1973.

\bibitem{Frederico:2010zh}
T. Frederico and G. Salm\`e.
\newblock {Projecting the Bethe-Salpeter Equation onto the Light-Front and
  back: A Short Review}.
\newblock {\em Few Body Syst.}, 49:163--175, 2011.

\bibitem{Glazek:1992bs}
S.~D. Glazek, A. Harindranath, S. Pinsky, J. Shigemitsu,
  and K. G. Wilson.
\newblock {On the relativistic bound state problem in the light front Yukawa
  model}.
\newblock {\em Phys. Rev. D}, 47:1599--1619, 1993.

\bibitem{Mangin-Brinet:2001wmj}
M.~Mangin-Brinet, J.~Carbonell, and V.~A. Karmanov.
\newblock {Relativistic bound states in Yukawa model}.
\newblock {\em Phys. Rev. D}, 64:125005, 2001.

\bibitem{ManginBrinet:2003nm}
M.~Mangin-Brinet, J.~Carbonell, and V.~A. Karmanov.
\newblock {Two fermion relativistic bound states in Light Front Dynamics}.
\newblock {\em Phys. Rev. C}, 68:055203, 2003.

\bibitem{Bakker:2006pn}
B. L.~G. Bakker, J.~K. Boomsma, and C.-R. Ji.
\newblock {The Box diagram in Yukawa theory}.
\newblock {\em Phys. Rev. D}, 75:065010, 2007.

\bibitem{Marinho:2007zzb}
J.~A.~O. Marinho, T.~Frederico, and P.~U. Sauer.
\newblock {Light-front Ward-Takahashi identity and current conservation}.
\newblock {\em Phys. Rev. D}, 76:096001, 2007.

\bibitem{Marinho:2008pe}
J.~A.~O. Marinho, T.~Frederico, E.~Pace, G.~Salm\`e, and P.~Sauer.
\newblock {Light-front Ward-Takahashi Identity for Two-Fermion Systems}.
\newblock {\em Phys. Rev. D}, 77:116010, 2008.

\bibitem{Vary:2016emi}
J.~P. Vary, L. Adhikari, G. Chen, Yang Li, P. Maris, and X.
  Zhao.
\newblock {Basis Light-Front Quantization: Recent Progress and Future Prospects}.
\newblock {\em Few Body Syst.}, 57(8):695--702, 2016.



\bibitem{LiuSoperPRD1993}
H.~H. Liu and D.~E. Soper.
\newblock Implementation of the leibbrandt-mandelstam gauge prescription in the
  null-plane bound-state equation.
\newblock {\em Phys. Rev. D}, 48:1841--1851, Aug 1993.

\bibitem{Nakanishi1977}
N.~Nakanishi and H.~Yabuki.
\newblock Null-plane quantization and haag's theorem.
\newblock {\em Letters in Mathematical Physics}, 1:371--374, 1977.

\bibitem{deMelo:1998an}
J.~P. B.~C de~Melo, J.~H.~O. Sales, T.~Frederico, and P.~U. Sauer.
\newblock {Pairs in the light front and covariance}.
\newblock {\em Nucl. Phys. A}, 631:574C--579C, 1998.

\bibitem{deMelo:2012hj}
J.~P. B.~C. de~Melo and T.~Frederico.
\newblock {Light-Front projection of spin-1 electromagnetic current and
  zero-modes}.
\newblock {\em Phys. Lett. B}, 708:87--92, 2012.

\bibitem{Collins:2018aqt}
J. Collins.
\newblock The non-triviality of the vacuum in light-front quantization: An
  elementary treatment.
\newblock arXiv:1801.03960 [hep-ph].

\bibitem{Chabysheva:2025vtm}
S.~S. Chabysheva and J.~R. Hiller.
\newblock Zero modes on the light front.
\newblock arXiv:2502.01775 [hep-th].



\bibitem{Moita:2022lfu}
R.~M. Moita, J.~P. B.~C. de~Melo, T.~Frederico, and W.~de~Paula.
\newblock {Pion inspired by QCD: Nakanishi and light-front integral
  representations}.
\newblock {\em Phys. Rev. D}, 106(1):016016, 2022.

\bibitem{Frederico:2003zk}
T.~Frederico, J.~H.~O. Sales, B.~V. Carlson, and P.~U. Sauer.
\newblock {Light front time picture of few body systems}.
\newblock {\em Nucl. Phys. A}, 737:260--264, 2004.

\bibitem{Pauli:1998tf}
H.~C. Pauli.
\newblock {On confinement in a light cone Hamiltonian for QCD}.
\newblock {\em Eur. Phys. J. C}, 7:289--303, 1999.

\bibitem{Pauli:1999hi}
H.-C. Pauli.
\newblock {Discretized light cone quantization and the effective interaction in
  hadrons}.
\newblock {\em AIP Conf. Proc.}, 494(1):80--139, 1999.

\bibitem{Marinho:2007zz}
J.~A.~O. Marinho and T.~Frederico.
\newblock {Next-to-leading order light-front three-body dynamics}.
\newblock {\em PoS}, LC2008:036, 2008.

\bibitem{Marinho:2008zza}
J.~A.~O. Marinho and T.~Frederico.
\newblock {Three-boson systems in light-front dynamics}.
\newblock {\em J. Phys. Conf. Ser.}, 110:122009, 2008.

\bibitem{Guimaraes:2014kor}
K.~S.~F.~F. Guimar\~aes, O.~Louren\c{c}o, W.~de~Paula, T.~Frederico, and A.~C.
  dos Reis.
\newblock {Final state interaction in $D^{+} \to K^{-} \pi^{+} \pi^{+}$ with
  $K\pi$ I = 1/2 and 3/2 channels}.
\newblock {\em JHEP}, 08:135, 2014.

\bibitem{Ydrefors:2021mky}
E. Ydrefors and T. Frederico.
\newblock {Proton image and momentum distributions from light-front dynamics}.
\newblock {\em Phys. Rev. D}, 104(11):114012, 2021.

\bibitem{Frederico:1992np}
T.~Frederico.
\newblock {Null plane model of three bosons with zero range interaction}.
\newblock {\em Phys. Lett. B}, 282:409--414, 1992.

\bibitem{Carbonell:2002qs}
J.~Carbonell and V.~A. Karmanov.
\newblock {Three boson relativistic bound states with zero range interaction}.
\newblock {\em Phys. Rev. C}, 67:037001, 2003.

\bibitem{deAraujo:1995mh}
W.~R.~B. de~Ara\'ujo, J.~P. B.~C. de~Melo, and T.~Frederico.
\newblock {Faddeev null plane model of the nucleon}.
\newblock {\em Phys. Rev. C}, 52:2733--2737, 1995.

\bibitem{Suisso:2002jg}
E.~F. Suisso, J.~P. B.~C. de~Melo, and T.~Frederico.
\newblock {Relativistic dynamics of Qqq systems}.
\newblock {\em Phys. Rev. D}, 65:094009, 2002.

\bibitem{Xu:2021wwj}
S. Xu, C. Mondal, J. Lan, X. Zhao, Y. Li, and J.~P.
  Vary.
\newblock {Nucleon structure from basis light-front quantization}.
\newblock {\em Phys. Rev. D}, 104(9):094036, 2021.

\bibitem{Xu:2024sjt}
S. Xu, Y. Liu, C. Mondal, J. Lan, X. Zhao, Y. Li, and
  J.~P. Vary.
\newblock {Towards a first principles light-front Hamiltonian for the nucleon}.
\newblock {\em Phys. Lett. B}, 867:139599, 2025.

\bibitem{Eichmann:2016yit}
G. Eichmann, H. Sanchis-Alepuz, R. Williams, R. Alkofer, and
  C.~S. Fischer.
\newblock {Baryons as relativistic three-quark bound states}.
\newblock {\em Prog. Part. Nucl. Phys.}, 91:1--100, 2016.

\bibitem{Segovia:2015hra}
J. Segovia, B. El-Bennich, E. Rojas, I.~C. Cloet, C.~D. Roberts,
  S.-S. Xu, and H.-S. Zong.
\newblock {Completing the picture of the Roper resonance}.
\newblock {\em Phys. Rev. Lett.}, 115(17):171801, 2015.

\bibitem{Eichmann:2018adq}
G. Eichmann and C.~S. Fischer.
\newblock {Baryon Structure and Reactions}.
\newblock {\em Few Body Syst.}, 60(1):2, 2019.

\bibitem{Liu:2023reo}
L. Liu and C.~S. Fischer.
\newblock {Space-like electromagnetic form factors of lambda- and sigma-baryons
  from quark-diquark Faddeev equations}.
\newblock {\em Eur. Phys. J. A}, 60(4):84, 2024.

\bibitem{Chen:2023zhh}
C. Chen, C.~S. Fischer, and C.~D. Roberts.
\newblock {Nucleon-to-\ensuremath{\Delta} Axial and Pseudoscalar Transition
  Form Factors}.
\newblock {\em Phys. Rev. Lett.}, 133(13):131901, 2024.

\bibitem{Yu:2024qsd}
Y. Yu, P. Cheng, H.-Y. Xing, F. Gao, and C.~D. Roberts.
\newblock {Contact interaction study of proton parton distributions}.
\newblock {\em Eur. Phys. J. C}, 84(7):739, 2024.

\bibitem{Oliveira:2018fkj}
O. Oliveira, T.~Frederico, W.~de~Paula, and J.~P. B.~C. de~Melo.
\newblock {Exploring the Quark-Gluon Vertex with Slavnov-Taylor Identities and
  Lattice Simulations}.
\newblock {\em Eur. Phys. J. C}, 78(7):553, 2018.


\bibitem{Ydrefors:2020duk}
E.~Ydrefors, J.~H. Alvarenga~Nogueira, V.~A. Karmanov, and T.~Frederico.
\newblock {Three-boson bound states in Minkowski space with contact
  interactions}.
\newblock {\em Phys. Rev. D}, 101(9):096018, 2020.

\bibitem{Magalhaes:2011sh}
P.~C. Magalh\~aes, M.~R. Robilotta, K.~S. F.~F. Guimar\~aes, T.~Frederico,
  W.~de~Paula, I.~Bediaga, A.~C.~dos Reis, C.~M. Maekawa, and G.~R.~S.
  Zarnauskas.
\newblock {Towards three-body unitarity in $D^+ \to K^- \pi^+ \pi^+$}.
\newblock {\em Phys. Rev. D}, 84:094001, 2011.

\bibitem{Bediaga:2024ipi}
I.~Bediaga, M.~A. Shalchi, T.~Frederico, and P.~C. Magalh\~aes.
\newblock {Hadronic rescattering to solve the helicity puzzle in $B^+\to p\bar
  p\pi^+(K^+)$ decays}.
\newblock {\em Phys. Rev. D}, 110(9):096026, 2024.

\bibitem{Sauli:2001we}
V. Sauli and J.~Adam, Jr.
\newblock {Study of relativistic bound states for scalar theories in the
  Bethe-Salpeter and Dyson-Schwinger formalism}.
\newblock {\em Phys. Rev. D}, 67:085007, 2003.

\bibitem{Carbonell:2005nw}
J.~Carbonell and V.~A. Karmanov.
\newblock {Cross-ladder effects in Bethe-Salpeter and light-front equations}.
\newblock {\em Eur. Phys. J. A}, 27:11--21, 2006.

\bibitem{Karmanov:2021imh}
V.~A. Karmanov.
\newblock {New form of kernel in equation for Nakanishi function}.
\newblock {\em Phys. Rev. D}, 104(5):056012, 2021.

\bibitem{Carbonell:2013kwa}
J.~Carbonell and V.~A. Karmanov.
\newblock {Bethe-Salpeter scattering amplitude in Minkowski space}.
\newblock {\em Phys. Lett. B}, 727:319--324, 2013.

\bibitem{Carbonell:2014dwa}
J.~Carbonell and V.~A. Karmanov.
\newblock {Solving Bethe-Salpeter scattering state equation in Minkowski
  space}.
\newblock {\em Phys. Rev. D}, 90(5):056002, 2014.

\bibitem{Eichmann:2021vnj}
G. Eichmann, E. Ferreira, and A. Stadler.
\newblock {Going to the light front with contour deformations}.
\newblock {\em Phys. Rev. D}, 105(3):034009, 2022.

\bibitem{Huber:2022nzs}
M.~Q. Huber, W.~J. Kern, and R. Alkofer.
\newblock {Analytic structure of three-point functions from contour
  deformations}.
\newblock {\em Phys. Rev. D}, 107(7):074026, 2023.



\bibitem{Gutierrez:2016ixt}
C.~Gutierrez, V.~Gigante, T.~Frederico, G.~Salm\`e, M.~Viviani, and L. Tomio.
\newblock {Bethe-Salpeter bound-state structure in Minkowski space}.
\newblock {\em Phys. Lett. B}, 759:131--137, 2016.

\bibitem{Schwarz:2004mv}
J.~H. Schwarz.
\newblock {The Generalized Stieltjes transform and its inverse}.
\newblock {\em J. Math. Phys.}, 46:013501, 2005.

\bibitem{Castro:2023bij}
A.~Castro, W.~de~Paula, T.~Frederico, and G.~Salm\`e.
\newblock {Exploring the $0^-$ bound state with dressed quarks in Minkowski
  space}.
\newblock {\em Phys. Lett. B}, 845:138159, 2023.

\bibitem{Noronha:2023hjg}
A.~Noronha, W.~de~Paula, J.~H. de~Alvarenga Nogueira, T.~Frederico, E.~Pace,
  and G.~Salm\`e.
\newblock {Chiral limit of a fermion-scalar (1/2)+ system in covariant gauges}.
\newblock {\em Phys. Rev. D}, 107(9):096019, 2023.

\bibitem{Sauli:2002tk}
V. Sauli.
\newblock {Minkowski solution of Dyson-Schwinger equations in momentum
  subtraction scheme}.
\newblock {\em JHEP}, 02:001, 2003.

\bibitem{Mezrag:2020iuo}
C. Mezrag and G. Salm\`e.
\newblock {Fermion and Photon gap-equations in Minkowski space within the
  Nakanishi Integral Representation method}.
\newblock {\em Eur. Phys. J. C}, 81(1):34, 2021.

\bibitem{Duarte:2022yur}
D.~C. Duarte, T. Frederico, W. de~Paula, and E. Ydrefors.
\newblock {Dynamical mass generation in Minkowski space at QCD scale}.
\newblock {\em Phys. Rev. D}, 105(11):114055, 2022.

\bibitem{Alkofer:2018guy}
R. Alkofer, A. Maas, W.~A. Mian, M. Mitter, J.
  Par\'\i{}s-L\'opez, J.~M. Pawlowski, and N. Wink.
\newblock {Bound state properties from the functional renormalization group}.
\newblock {\em Phys. Rev. D}, 99(5):054029, 2019.

\bibitem{Dupuis:2020fhh}
N.~Dupuis, L.~Canet, A.~Eichhorn, W.~Metzner, J.~M. Pawlowski, M.~Tissier, and
  N.~Wschebor.
\newblock {The nonperturbative functional renormalization group and its
  applications}.
\newblock {\em Phys. Rept.}, 910:1--114, 2021.

\bibitem{AlvarengaNogueira:2017cpt}
J.~H. Alvarenga~Nogueira, C.-R. Ji, E.~Ydrefors, and T.~Frederico.
\newblock {Color-suppression of non-planar diagrams in bosonic bound states}.
\newblock {\em Phys. Lett. B}, 777:207--211, 2018.

\bibitem{Mezrag:2023nkp}
C. Mezrag.
\newblock {Generalised Parton Distributions in Continuum Schwinger Methods:
  Progresses, Opportunities and Challenges}.
\newblock {\em Particles}, 6(1):262--296, 2023.

\bibitem{Ydrefors:2022bhq}
E. Ydrefors and T. Frederico.
\newblock {Proton quark distributions from a light-front Faddeev-Bethe-Salpeter
  approach}.
\newblock {\em Phys. Lett. B}, 838:137732, 2023.

\bibitem{Barabanov:2020jvn}
M.~Yu. Barabanov et~al.
\newblock {Diquark correlations in hadron physics: Origin, impact and
  evidence}.
\newblock {\em Prog. Part. Nucl. Phys.}, 116:103835, 2021.

\bibitem{PDG2024}
S.~Navas et~al.
\newblock Review of $\text{P}$article $\text{P}$hysics.
\newblock {\em Phys. Rev. D}, 110(3):030001, 2024.

\bibitem{deMelo:1997hh}
J.~P. B.~C de~Melo and T.~Frederico.
\newblock {Covariant and light front approaches to the rho meson
  electromagnetic form-factors}.
\newblock {\em Phys. Rev. C}, 55:2043, 1997.

\bibitem{deMelo:1999gn}
J.~P. B.~C de~Melo, T.~Frederico, H.~W.~L. Naus, and P.~U. Sauer.
\newblock {Covariance of light front models: Pair current}.
\newblock {\em Nucl. Phys. A}, 660:219--231, 1999.

\bibitem{deMelo:2003cb}
J.~P. B.~C. de~Melo and T.~Frederico.
\newblock {Spin-1 particle in the light-front approach}.
\newblock {\em Braz. J. Phys.}, 34:881--884, 2004.

\bibitem{deMelo:2016lwr}
J.~P. B.~C. de~Melo, Chueng-Ryong Ji, and T.~Frederico.
\newblock {The \ensuremath{\rho} -meson time-like form factors in sub-leading
  pQCD}.
\newblock {\em Phys. Lett. B}, 763:87--93, 2016.

\bibitem{Lepage:1980fj}
G.~P. Lepage and S.~J. Brodsky.
\newblock {Exclusive Processes in Perturbative Quantum Chromodynamics}.
\newblock {\em Phys. Rev. D}, 22:2157, 1980.

\bibitem{Oliveira:2020yac}
O. Oliveira, T. Frederico, and W. de~Paula.
\newblock {The soft-gluon limit and the infrared enhancement of the quark-gluon
  vertex}.
\newblock {\em Eur. Phys. J. C}, 80(5):484, 2020.

\bibitem{Oliveira:2018ukh}
O. Oliveira, W. de~Paula, T. Frederico, and J.~P. B.~C. de~Melo.
\newblock {The Quark-Gluon Vertex and the QCD Infrared Dynamics}.
\newblock {\em Eur. Phys. J. C}, 79(2):116, 2019.

\bibitem{Zhu:2023lst}
Z. Zhu, Z. Hu, J. Lan, C. Mondal, X. Zhao, and J.~P.
  Vary.
\newblock {Transverse structure of the pion beyond leading twist with basis
  light-front quantization}.
\newblock {\em Phys. Lett. B}, 839:137808, 2023.

\bibitem{Pasquini:2023aaf}
B. Pasquini, S. Rodini, and S. Venturini.
\newblock {Valence quark, sea, and gluon content of the pion from the parton
  distribution functions and the electromagnetic form factor}.
\newblock {\em Phys. Rev. D}, 107(11):114023, 2023.

\bibitem{Lin:2024ijo}
B. Lin, S. Nair, C. Mondal, S. Xu, Z. Hu, P. Zhang,
  X. Zhao, and J.~P. Vary.
\newblock {Chiral-odd gluon generalized parton distributions in the proton: A
  light-front quantization approach}.
\newblock {\em Phys. Lett. B}, 860:139153, 2025.

\bibitem{Roberts:2023lap}
C.~D. Roberts.
\newblock {Hadron Structure Using Continuum Schwinger Function Methods}.
\newblock {\em Few Body Syst.}, 64(3):51, 2023.

\bibitem{LatticePartonCollaborationLPC:2022myp}
J.-C. He, M.-H. Chu, J. Hua, X. Ji, A. Sch\"afer, Y. Su,
  W. Wang, Y.-B. Yang, J.-H. Zhang, and Q.-A. Zhang.
\newblock {Unpolarized transverse momentum dependent parton distributions of the nucleon from lattice QCD}.
\newblock {\em Phys. Rev. D}, 109(11):114513, 2024.

\bibitem{LatticeParton:2023xdl}
M.-H. Chu et~al.
\newblock {Transverse-momentum-dependent wave functions of the pion from
  lattice QCD}.
\newblock {\em Phys. Rev. D}, 109(9):L091503, 2024.

\bibitem{FLAG:2024oxs}
Y.~Aoki \textit{et al.} [Flavour Lattice Averaging Group (FLAG)],
``FLAG Review 2024,''
[arXiv:2411.04268 [hep-lat]].



\end{thebibliography}

\end{document}